\documentstyle[prl,aps,floats,preprint,epsfig]{revtex}

\begin{document}

\renewcommand{\topfraction}{1.0}
\renewcommand{\bottomfraction}{1.0}
\renewcommand{\textfraction}{0.0}

\def\Uone  {\relax\ifmmode{\Upsilon(1S)}
                 \else{$\Upsilon(1S)$}\fi}
\def\Utwo  {\relax\ifmmode{\Upsilon(2S)}
                 \else{$\Upsilon(2S)$}\fi}
\def\Uthre  {\relax\ifmmode{\Upsilon(3S)}
                 \else{$\Upsilon(3S)$}\fi}
\def\Ufour {\relax\ifmmode{\Upsilon(4S)}
                 \else{$\Upsilon(4S)$}\fi}
\def\qq {\relax\ifmmode{q\overline{q}}
                \else{$q\overline{q}$}\fi}
 
\newcommand{\decpiall}{\mbox{$\Utwo\rightarrow \Uone\pi\pi$}}
\newcommand{\decone}{\mbox{$\Utwo\rightarrow \Uone\pi^+\pi^-$}}
\newcommand{\dectwo}{\mbox{$\Upsilon(2S)\rightarrow \Upsilon(1S)\pi^0\pi^0$}}
\newcommand{\decthree}{\mbox{$\Upsilon(2S)\rightarrow \Upsilon(1S)\eta$}}
\newcommand{\decfour}{\mbox{$\Upsilon(2S)\rightarrow \Upsilon(1S)\pi^0$}}
\newcommand{\decsevn}{\mbox{$\Upsilon(1S)\rightarrow X$}}
\newcommand{\deceigh}{\mbox{$\Upsilon(1S)\rightarrow l^+l^-$}}
\newcommand{\decnine}{\mbox{$\eta\rightarrow \gamma\gamma$}}
\newcommand{\decten}{\mbox{$\eta\rightarrow 3\pi^0$}}
\newcommand{\decelev}{\mbox{$\eta\rightarrow \pi^+\pi^-\pi^0$}}
\newcommand{\dectwel}{\mbox{$\eta\rightarrow \pi^+\pi^-\gamma$}}

\newcommand{\chainone}{\decone,\decfive}
\newcommand{\chaintwo}{\decone,\decsix}

\def\Nincl {\relax\ifmmode{ 50566\pm  570 }
       \else{$50566\pm  570 $}\fi} 
\def\Etaincl {\relax\ifmmode{ 52.9\pm  2.0 }
       \else{$52.9\pm  2.0 $}\fi} 
\def\Nfive {\relax\ifmmode{ 0.196\pm 0.002\pm 0.010 }
       \else{$ 0.196\pm 0.002\pm 0.010 $}\fi} 

\def\BRone {\relax\ifmmode{ 0.189\pm 0.004\pm 0.010 }
         \else{$ 0.189\pm 0.004\pm 0.010  $}\fi} 
\def\Nbkgrelone {$  8.7$}
\def\Nbkgrmuone {$  3.8 $}
\def\Nbkgrprelone {$  0.9 $}
\def\Nbkgrprmuone {$ 0.3 $}
\def\dNbkgrelone {$  0.3 $}
\def\dNbkgrmuone {$  0.2 $}

\def\Nsix  {\relax\ifmmode{0.0229\pm 0.0008 \pm 0.0011}
       \else{$0.0229\pm 0.0008\pm 0.0011$}\fi}  
\def\Nsevn  {\relax\ifmmode{0.0249\pm 0.0008 \pm 0.0013}
       \else{$0.0249\pm 0.0008\pm 0.0013$}\fi}  

\def\Broneave { \relax\ifmmode{0.192\pm 0.002\pm 0.010}
       \else{$ 0.192\pm 0.002\pm 0.010 $}\fi } 

\def\BRtwo {\relax\ifmmode{ 0.092\pm 0.006\pm 0.008 }
       \else{$ 0.092\pm 0.006\pm 0.008  $}\fi} 
\def\Nbkgreltwo {$ 3.8 $}
\def\Nbkgrmutwo {$  1.4 $}
\def\Nbkgrpreltwo {$  2.9 $}
\def\Nbkgrprmutwo {$  1.0 $}
\def\dNbkgreltwo {$  1.5 $}
\def\dNbkgrmutwo {$  0.9 $}

\def\invmassdf { 13 }

\def\Epstwopitwol  { 0.077 \pm 0.041 }     

\def\Epstwopi  { 0.028\pm 0.027  }     

\def\Epstwopicomb { 0.042\pm 0.022 }     
\def\CLEpszero {  40.2 }

\def\Nthre  { 0.0028 }                
\def\Nsgnltot {  6 }

\def\Lumtwoson {$  73.57 $}
\def\Lumtwosoff {$   5.17 $}
\def\Nbkgtot {$   1.1 $}
\def\Neebkgpizerotwo {$   0.2 $}
\def\Nmmbkgpizerotwo {$   0.2 $}
\def\Neebkgpifour {$   0.3 $}
\def\Nmmbkgpifour {$   0.6 $}
\def\Neebkgtottwo {$  14.5 $}
\def\Nmmbkgtottwo {$   0.2 $}
\def\Neebkgtotfour {$   0.3 $}
\def\Nmmbkgtotfour {$   0.6 $}
\def\Neebkgconttwo {$  14.2 $}
\def\Nmmbkgconttwo {$   0.0 $}

\def\Nfour  { 0.0009 }                  
\def\Etapizel  {$ 29.3\pm  0.8 $ }      
\def\Etapizmu  {$ 36.3\pm  0.9 $ }      
\def\Npizeltot {  9 }
\def\Npizmutot {  6 }
\def\Npiztot { 15 }
\def\Npizbkgtot { 18.7 }
\def\Npizsig {  7  }

\def\Npizbkgtotgsb { 12.9 }
\def\Npizsiggsb {  8  }
\def\Brpizgsb { 0.0011 }

\preprint{\tighten \vbox{ \hbox{\hfil CLNS 98/1540}\hbox{\hfil CLEO 98-1}}
}

\title{The hadronic transitions \boldmath{$\Utwo\rightarrow \Uone$} \\}
\author{CLEO Collaboration}
\date{\today}

\maketitle
\tighten

\begin{abstract}
Using a 73.6 pb$^{-1}$ data sample of \Utwo\ events collected with the CLEO II detector at the Cornell Electron Storage Ring, we have investigated the hadronic transitions between the \Utwo\ and the \Uone. The dipion transition \decone\ was studied using two different analysis techniques. Selecting events in which $\Uone \rightarrow e^+e^-,\mu^+\mu^-$ (``exclusive'' analysis), and using the \Uone\ leptonic branching ratios world averages from the PDG review, we found ${\cal B}(\decone)=\BRone$, while using a method allowing $\Uone \rightarrow anything$ (``inclusive'' analysis) we found ${\cal B}(\decone)=\Nfive$.  The appropriate average of the two measurements gives ${\cal B}(\decone)=\Broneave$.  Combining the exclusive and inclusive results we derive the \Uone\ leptonic branching ratios ${\cal B}_{ee}=\Nsix$\ and ${\cal B}_{\mu\mu}=\Nsevn$. We also studied \dectwo\  and obtained ${\cal B}(\Upsilon(2S)\rightarrow \Upsilon(1S)\pi^0\pi^0)=\BRtwo$. Parameters of the $\pi\pi$ system (dipion invariant mass spectra, angular distributions) were analyzed and found to be consistent with current theoretical models. Lastly, we searched for the $\eta$ and single $\pi^0$ transitions and  obtained the upper limits ${\cal B}(\decthree) < \Nthre$ and ${\cal B}(\Upsilon(2S)\rightarrow \Upsilon(1S)\pi^0) < \Brpizgsb$. 
\end{abstract}


\newpage

{
\renewcommand{\thefootnote}{\fnsymbol{footnote}}

\begin{center}
J.~P.~Alexander,$^{1}$ R.~Baker,$^{1}$ C.~Bebek,$^{1}$
B.~E.~Berger,$^{1}$ K.~Berkelman,$^{1}$ K.~Bloom,$^{1}$
V.~Boisvert,$^{1}$ D.~G.~Cassel,$^{1}$ D.~S.~Crowcroft,$^{1}$
M.~Dickson,$^{1}$
% ...
\end{center}
 
\small
\begin{center}
$^{1}${Cornell University, Ithaca, New York 14853}\\
% ...
$^{13}${University of Oklahoma, Norman, Oklahoma 73019}\\
% ...
$^{20}${Virginia Polytechnic Institute and State University,
Blacksburg, Virginia 24061}\\
% ...
$^{25}${University of Colorado, Boulder, Colorado 80309-0390}
\end{center}


\setcounter{footnote}{0}
}
\newpage

\section{Introduction}

The hadronic transitions in heavy quarkonium systems provide an experimental testing ground for the theoretical calculations of nonperturbative QCD~\cite{qcdcalculations} and can give information on the structure of QCD confinement as well as on the gluon content of light hadrons. Historically, studies of the hadronic transitions \decpiall\ were preceded by investigations of the transitions $\eta' \rightarrow \eta\pi\pi$ and $\psi' \rightarrow \psi\pi\pi$. All three are examples of $\Delta I=0$ dipion transitions. In the decay $\eta' \rightarrow \eta\pi\pi$ the pions fit reasonably well to a phase space mass spectrum~\cite{etadipion}. Soon after the discovery of charmonium~\cite{psidipion}, and the subsequent observation of the $\psi' \rightarrow \psi\pi\pi$ transition, it was found that in this transition the dipion invariant mass spectrum cannot be adequately described by a phase space mass spectrum. The challenge of providing an acceptable description of the observed data attracted considerable theoretical attention. With the discovery of another family of heavy quarkonium states, the family of $\Upsilon$ resonances, the theoretical calculations were extended to include bottomonium.   

\begin{figure}[hbt]
\center
\epsfig{file=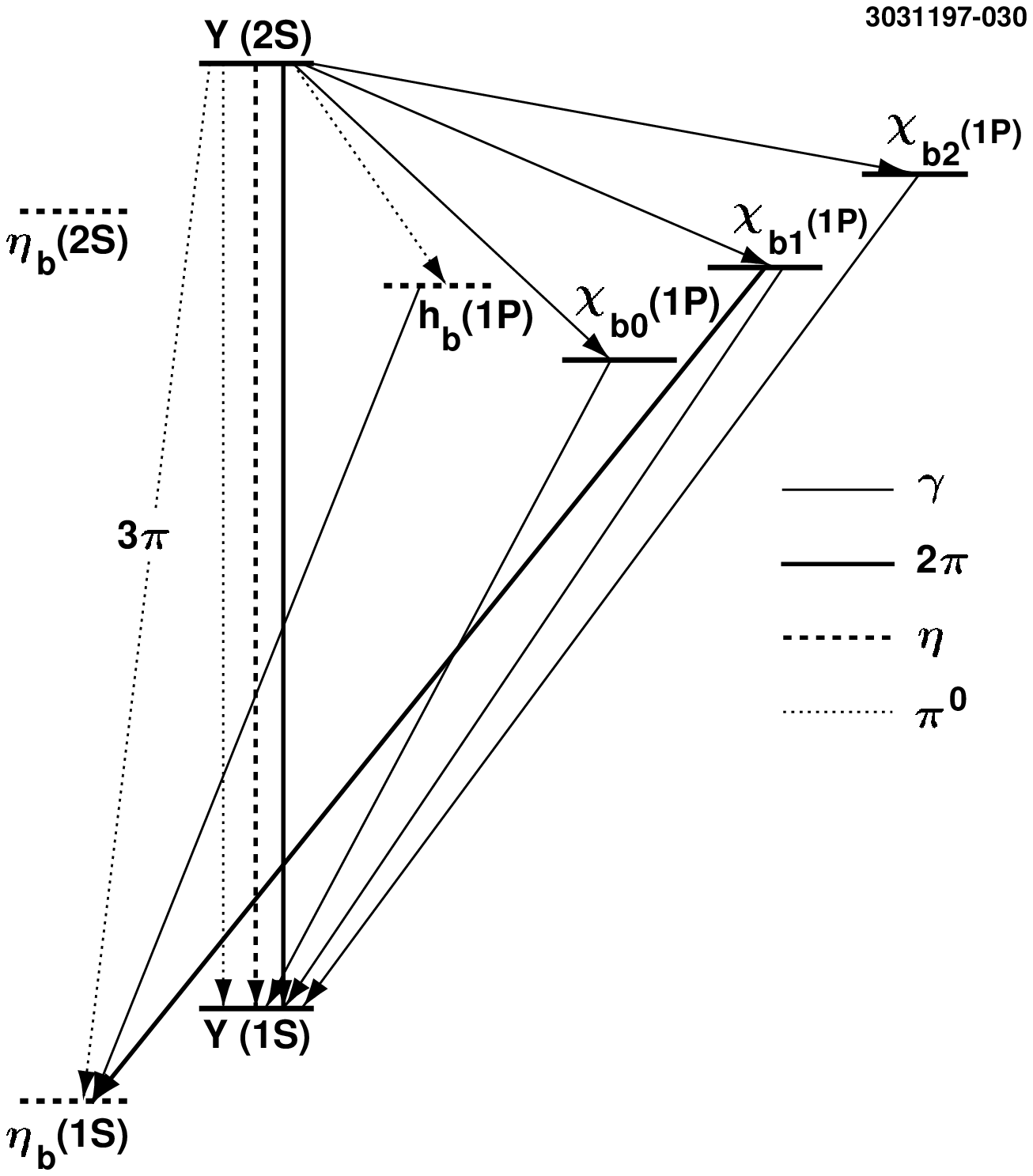, height=7cm}
\caption[]{Transitions in the bottomonium.}\label{figs:botlevels}
\end{figure}

Figure~\ref{figs:botlevels} shows the bottomonium levels up to the \Utwo\ and possible transitions between them, including radiative and rare ($3\pi$ and single $\pi^0$) transitions~\cite{REVIEW}. The hadronic transitions between the bottomonium levels are soft processes (typical transition energies are 0.3-0.9 GeV) and are thereby difficult to treat perturbatively. Typically, the heavy quarkonium hadronic transition $(\qq)' \rightarrow (\qq) X$ is treated as the factorizable product of two processes: first, the transition from $(\qq)'$ to (\qq)\ with the emission of gluons (usually two), followed by the hadronization of the gluons to the state $X$ (i.e., the production of $X$ from the vacuum in the presence of the gluon color field).

Although nonperturbative, the hadronic transitions between heavy quarkonia can nevertheless be described in the context of a ``multipole'' expansion scheme where the gluon fields are expanded in a multipole series, similar to the electromagnetic transitions, as first outlined by Gottfried~\cite{Gottfried}. In the framework of the multipole expansion, Yan~\cite{Yan}, and later Zhou and Kuang~\cite{Zhou-Kuang} calculated the transition rates and derived a parameterization for the dipion invariant mass spectrum in the \decpiall\ transitions. They used the quark-confining string model~\cite{quarkstring} to describe the intermediate state of the hadronic transition and calculate the hadronization matrix element. Rather then writing the gluonic degrees of freedom for the quark-confining string, Voloshin and Zakharov~\cite{Vol-Zakh} (VZ), and afterwards in a revised analysis Novikov and Shifman~\cite{Nov-Shif} (NS), used an alternate approach and wrote the general form of the QCD field tensor in the chiral limit to obtain the hadronization matrix element. In both approaches the hadronization matrix element is constrained by current algebra, partial conservation of the axial current (PCAC), and gauge invariance. The essential mass dependence of the matrix element is very similar in all cases: it vanishes for dipion mass  approaching threshold, and peaks at larger values of $m_{\pi\pi}$. In the NS and VZ models, the model parameters are derived from ``first-principles'', as opposed to the Yan {\em et al.} model where the parameters are determined phenomenologically from a fit to $\psi' \rightarrow \psi\pi\pi$. 

The results presented in this paper were obtained using the world's largest available data sample of \Utwo\ decays (73.6 pb$^{-1}$ of integrated luminosity on-resonance, and 5.2 pb$^{-1}$ off-resonance) collected with the CLEO II detector at the Cornell Electron Storage Ring operating at the \Utwo\ center of mass energy in December 1994. Similar investigations were performed by several collaborations including ARGUS~\cite{ARGUS87}, CUSB~\cite{CUSB84}, CLEO~\cite{CLEO84} and Crystal Ball~\cite{CBALL85}. Our data sample is larger by at least a factor of two in integrated luminosity than each of the previous measurements, with the number of \Utwo\ resonant decays $N_{\Utwo}=(488\pm 18)\cdot 10^3$ ~\cite{numberofutwos}.

\section{Detector}

CLEO II is a general purpose detector~\cite{cleodetector} for measuring charged and neutral particles in the energy range from $\approx 50$ MeV to $\approx 6$ GeV. Its three concentric wire drift chambers, covering 95\% of the solid angle, detect charged particles and perform particle identification using specific ionization energy loss measurements ($dE/dx$) in the outer chamber. A superconducting coil provides a magnetic field of 1.5 Tesla, giving a momentum resolution of $(\delta p/p)^2=(0.0015p)^2 + (0.005)^2$, where $p$ is the momentum in GeV/$c$. A time-of-flight system, just outside the drift chambers, consists of plastic scintillation counters and serves as a primary triggering system; it also provides some particle identification information. Beyond the time-of-flight system, but inside the solenoid, is an electromagnetic calorimeter, consisting of 7800 thallium-doped CsI crystals arranged as two endcaps and a barrel region. The central barrel region of the calorimeter covers 75\% of the solid angle and achieves an energy resolution of $\delta E/E(\%)=0.35/E^{0.75} + 1.9 -0.1E$, where $E$ is the shower energy in GeV. The endcaps of the calorimeter extend the solid angle coverage to about 95\% of $4\pi$, although energy resolution is not quite as good as in the barrel. Proportional tracking chambers for muon detection are located in between and outside of the iron slabs that provide the magnetic field flux return.

In our analysis we used a customized version of LUND/JETSET~\cite{jetset} program as a Monte Carlo event generator. The final state particles are then propagated through and decayed in the CLEO II detector using a GEANT~\cite{geant} based detector simulation.

\section{Transition \boldmath{\decone}}

We studied the dipion transitions \decpiall\ using two different techniques. The first one selects events where there is a\ $e^+e^-$ or $\mu^+\mu^-$ pair recoiling against the dipion system, which is assumed to result from  $\Uone\rightarrow e^+e^-,\mu^+\mu^-$ (``exclusive'' measurement). In the second technique we do not specify how the \Uone\ decays, selecting all events which have  a $\pi^+\pi^-$ pair (``inclusive'' measurement). The two measurements are complementary to each other and provide us with important cross-checks.

\subsection{\em Exclusive final states with $\Upsilon(1S)\rightarrow e^+e^-,\mu^+\mu^-$}

We use the following selection criteria for the exclusive events with $\pi^+\pi^-l^+l^-$ in the final state. We demand four tracks in the event which pass track quality requirements, two of them (the lepton candidate tracks) must have momentum greater than 3.5 GeV/$c$ and come from the vicinity of the interaction point which is defined as a cylinder of 3 mm $\times$ 10 cm (radius $\times$ length) with its axis lying along the beam direction, and the other two (pion candidate tracks) must have momentum less than 0.5 GeV/$c$ and come from a similar cylindrical region 4 mm $\times$ 12 cm around the interaction point. To suppress background from radiative Bhabha events with $\gamma$-conversion we require that the cosine of the angle between the pion tracks satisfy $\cos\theta_{\pi\pi}<0.9$. We identify  electrons by the combined requirement that the ratio of the shower energy to the momentum of the matching track is close to 1 and that the lateral energy deposition in the calorimeter is consistent with the electron hypothesis. Events with muons are identified by requiring that the sum of the maximum penetration depths of the two tracks into the muon system absorber be greater than four hadronic absorption lengths.\footnote{As one can notice we just require for two high momentum lepton tracks to be present in the event and do not apply any further criteria on the dilepton system.} 

\begin{figure}[hbt]
\center
\epsfig{file=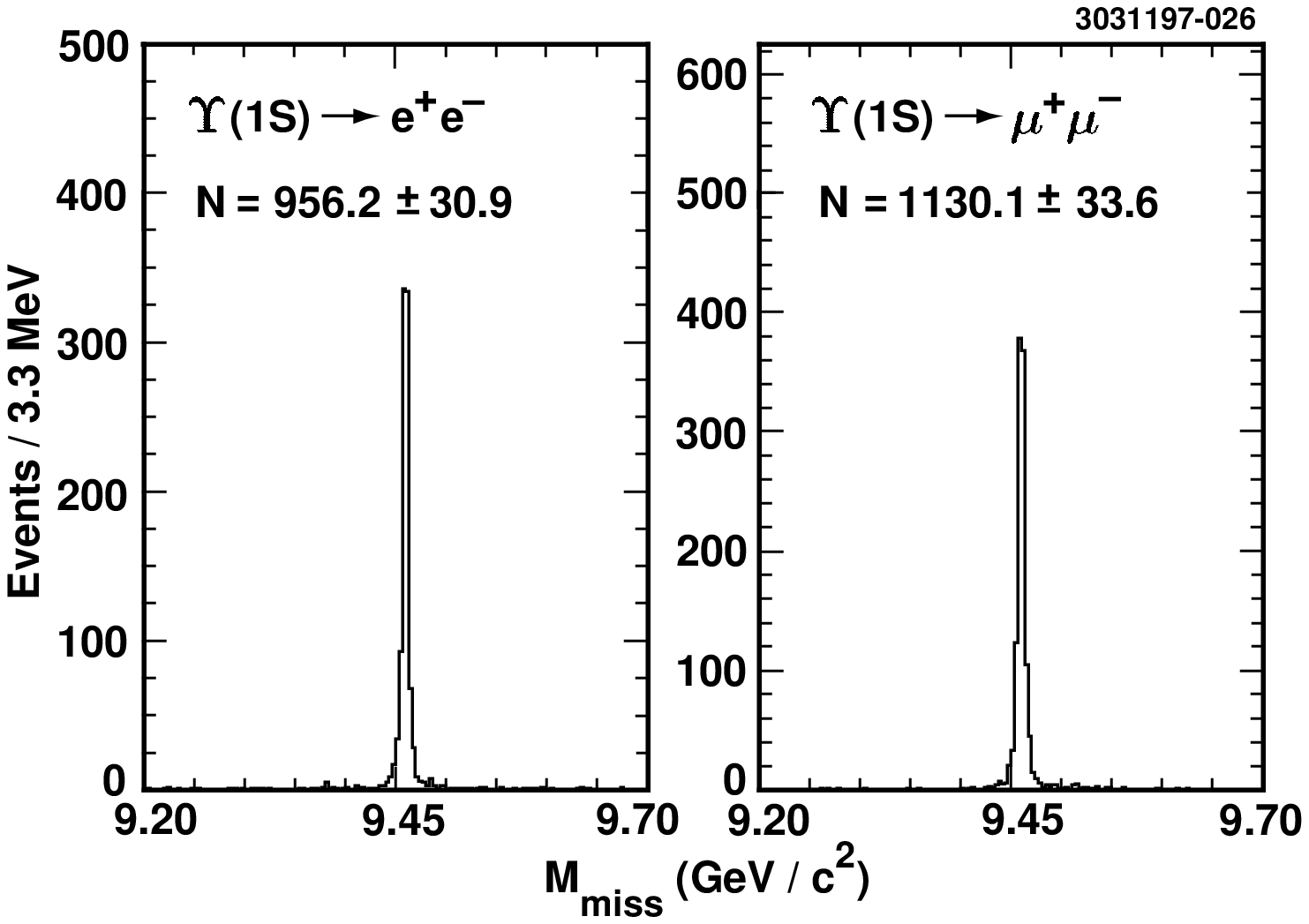, height=8cm, width=14cm}
\caption[]{The missing mass distributions in the exclusive $\Utwo\rightarrow \Uone\pi^+\pi^-$ measurement.}\label{figs:excl2pifit}
\end{figure}

The missing mass $M_{miss}=\sqrt{(M_{\Utwo}-E_{\pi\pi})^2-p_{\pi\pi}^2}$ (i.e., the mass recoiling against the dipion system) distributions for both $ee$ and $\mu\mu$ channels are shown in Figure~\ref{figs:excl2pifit}. One can see a clean signal with very little background in the side-bands,\footnote{The signal region is defined as the missing mass interval (9.43,9.49) GeV, the side-bands are defined as $(9.20,9.40)\cup (9.52,9.70)$ GeV in both dilepton channels.} thus we use a simple event count to obtain the number of observed events both in Monte Carlo (to calculate efficiencies) and in data.

The three largest sources of background are QED radiative processes with $\gamma$-conversion, two-photon double-tag production of $\pi\pi$ (in the $ee$ channel) and one-prong $\tau$-decays from $\Uone\rightarrow \tau\tau$. Due to our high lepton momentum and lepton identification requirements the contamination from $\tau$-decays to our data sample (which we directly subtract from the number of observed events) is very small: less then one event in each channel considered. To eliminate QED radiative and two-photon background we use the method of side-band subtraction: we count the number of events in the side-bands of our signal region and extrapolate this number into the signal region. In this way, we find the background contamination to be \Nbkgrelone\ events (\Nbkgrprelone\%)\ in the $ee$ channel and \Nbkgrmuone\ events (\Nbkgrprmuone\%)\ in the $\mu\mu$ channel.

\begin{table}[h]
\center
\caption[]{\small Numbers of events observed after background subtraction, efficiencies, product of branching ratios ${\cal B}(\decone)\cdot {\cal B}(\deceigh)$ and branching ratio ${\cal B}(\decone)$ for the exclusive measurement.}\label{tab:excl2piratio}
\begin{tabular}{ccccc}
 & $N^{observed}$  & $\epsilon$ (\%) & ${\cal B}_{\pi\pi}\cdot {\cal B}_{ll}\ (\times 10^{-3})$ & ${\cal B}_{\pi\pi}$ \\ \hline
 $ee$ & $  956.2\pm  30.9 $&$ 43.7\pm 1.4 $&$ 4.5\pm 0.1\pm 0.2 $&$ 0.178\pm 0.006\pm 0.015 $ \\
 $\mu\mu$ & $ 1130.1\pm  33.6 $&$ 47.5\pm 1.6 $&$ 4.9\pm 0.1\pm 0.2 $&$ 0.196\pm 0.006\pm 0.011 $ \\
\end{tabular}
\end{table}

Knowing the efficiencies $\epsilon_{ll}$ from the Monte Carlo simulation,\footnote{For all our sub-analyses we used the Voloshin and Zakharov~\cite{Vol-Zakh} model with $\lambda=3.44$ to generate the dipion invariant mass spectrum in the Monte Carlo simulation.} we can calculate the products of two branching ratios ${\cal B}(\Utwo\rightarrow \Uone\pi^+\pi^-)\cdot {\cal B}(\deceigh)=N^{observed}_{ll}/(\epsilon_{ll}N_{\Utwo})$, as shown in Table~\ref{tab:excl2piratio}. Using the Particle Data Group (PDG) values~\cite{pdgleptonbr} for ${\cal B}(\Uone\rightarrow e^+e^-)=0.0252\pm 0.0017$ and ${\cal B}(\Uone\rightarrow \mu^+\mu^-)=0.0248\pm 0.0007$ we obtain the \Utwo\ dipion branching fraction. Combining the results from both channels we find:
\[{\cal B}(\decone)=\BRone \]
where the first error is statistical and the second is systematic\footnote{When we average over the two dilepton channels, we treat correlated and uncorrelated errors separately in calculating the overall systematic error.} (for a breakdown of systematic errors see Sec.V).

In Table~\ref{tab:exclcomp} we compare our result with other exclusive measurements.

\begin{table}[ht]
\center
\caption[]{\small ${\cal B}(\decone)\cdot {\cal B}(\deceigh)$ in units of $10^{-3}$.}\label{tab:exclcomp}
\begin{tabular}{ll}
ARGUS~\cite{ARGUS87}          & $4.4\pm 0.2\pm 0.4$ \\
Crystal Ball~\cite{CBALL85}   & $4.9\pm 0.4\pm 1.0$ \\
CUSB~\cite{CUSB84}            & $5.4\pm 0.3\pm 0.4$ \\
CLEO~\cite{CLEO84}            & $5.4\pm 0.4$ \\
LENA~\cite{LENA81}            & $6.1\pm 2.3$ \\
 this analysis & $ 4.66\pm 0.10\pm 0.23 $ \\ \hline
 average         & $ 4.82 \pm 0.18 $ \\
\end{tabular}
\end{table}

\subsection{\em Inclusive final states}

In our inclusive analysis of \decone,\decsevn\ we select events with at least two tracks which pass our track quality requirements, have momentum less than 0.5 GeV/$c$ and come from the interaction region. We also require that the invariant mass of the two pion candidates lie between 0.27 GeV/$c^2$ and 0.57 GeV/$c^2$.

\begin{figure}[htb]
\center
\epsfig{file=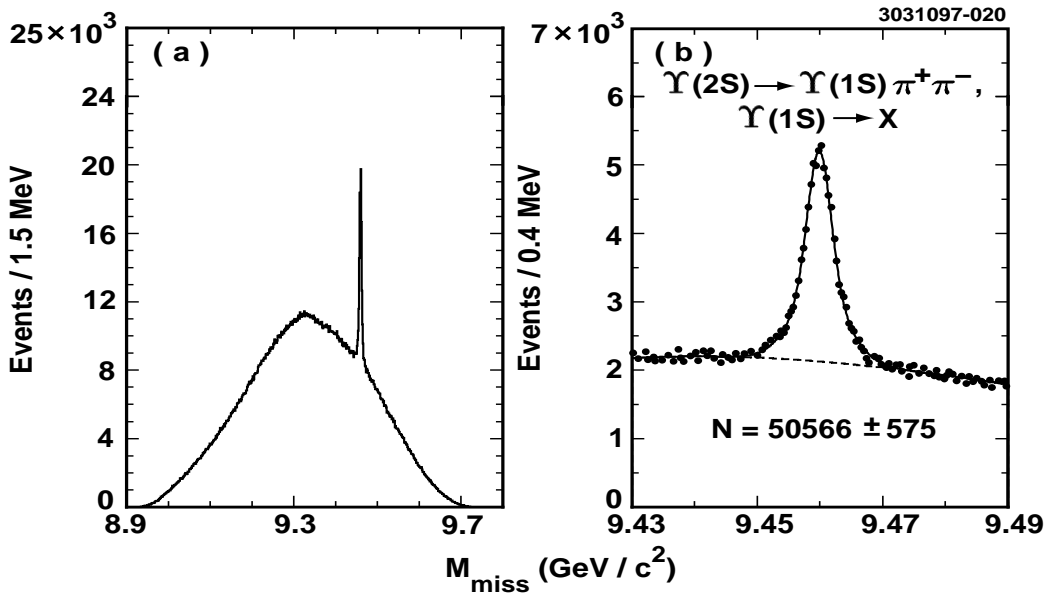, height=8cm, width=14cm}
\caption[]{Missing mass distribution from the inclusive $\Utwo\rightarrow \Uone\pi^+\pi^-$ events: a) the full distribution, b) the region near the \Uone\ mass, with the fit to the \Uone\ peak.}\label{figs:2pi-fit}
\end{figure}

The signal appears in the missing mass plot shown in Figure~\ref{figs:2pi-fit} along with the fit to the \Uone\ peak. The fitting function we use is a double-gaussian with the difference between the means fixed at zero, plus a third order polynomial for the background. The number of fitted events in the peak  is $N_{incl}=\Nincl$. The efficiency for the \decone\ channel has been calculated from a Monte Carlo simulation and determined to be $\epsilon_{incl}=(\Etaincl)\%$. From these two numbers and the total number of \Utwo\ produced we find the branching ratio for the transition \decone:
\[ {\cal B}(\decone)=\frac{N_{incl}}{\epsilon_{incl}N_{\Utwo}}=\Nfive \]

A comparison of this result with previous inclusive measurements is given in Table~\ref{tab:inclcomp}.
\begin{table}[ht]
\center
\caption[]{\small ${\cal B}(\decone)$ inclusive measurements.}\label{tab:inclcomp}
\begin{tabular}{ll} 
LENA~\cite{LENA81}   & $0.26\pm 0.13$ \\
ARGUS~\cite{ARGUS87} & $0.181\pm 0.005\pm 0.010$ \\
CLEO~\cite{CLEO84}   & $0.191\pm 0.012\pm 0.006$ \\
 this analysis & $ 0.196\pm 0.002\pm 0.010$\\ \hline
 average         & $ 0.190 \pm 0.007 $ \\
\end{tabular}
\end{table}

Combining the results of the exclusive and inclusive measurements, and taking into account correlations between the systematic errors, we obtain:\footnote{This number is subject to the use of the PDG values for the \Uone\ leptonic branching ratios.}
\[ {\cal B}(\decone)=\Broneave \]

Knowing the number of inclusive and exclusive events, we can now calculate the \Uone\ leptonic branching ratios ${\cal B}(\deceigh)=(N_{ll}\epsilon_{incl})/(N_{incl}\epsilon_{ll})$:
\[ B_{ee}={\cal B}(\Uone\rightarrow e^+e^-)=\Nsix \]
\[ B_{\mu\mu}={\cal B}(\Uone\rightarrow \mu^+\mu^-)=\Nsevn \]
which agree well with the corresponding PDG values.

\section{Transition \boldmath{\dectwo}}

To analyze the transition \dectwo\ exclusively in the final states with $\Upsilon(1S)\rightarrow e^+e^-,\mu^+\mu^-$, we apply criteria similar to those used in our $\Utwo\rightarrow \Uone\pi^+\pi^-$ exclusive analysis: we require two good quality charged tracks passing the criteria for a lepton candidate, and the same $e/\mu$-identification criteria.

\begin{figure}[htb]
\center
\epsfig{file=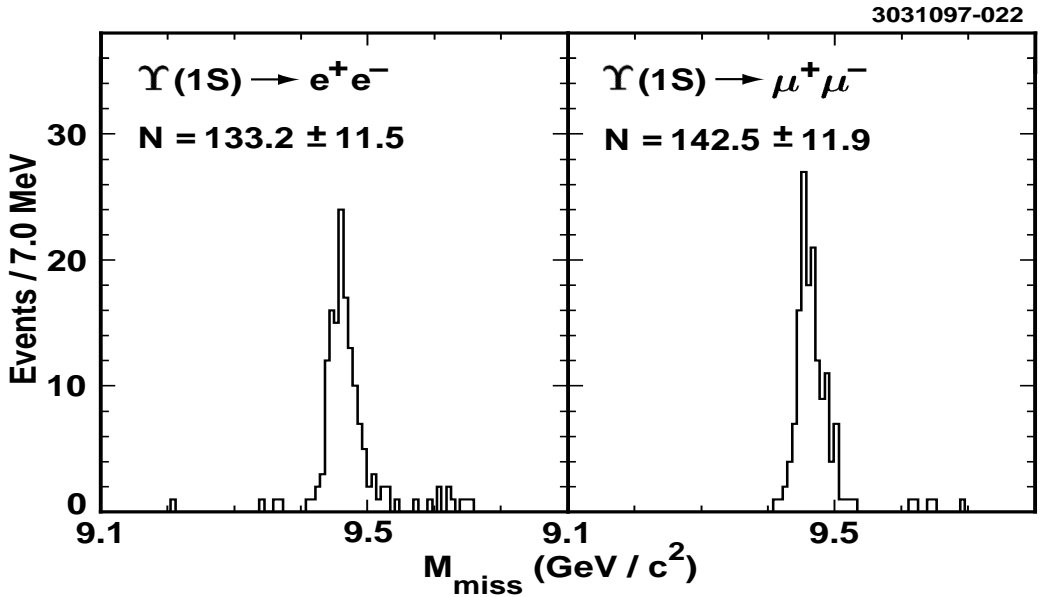, height=8cm, width=14cm}
\caption[]{The missing mass distributions in the exclusive $\Utwo\rightarrow \Uone\pi^0\pi^0$ measurement.}\label{figs:excl2pi0fit}
\end{figure}

We reconstruct $\pi^0$ candidates from photon showers in the calorimeter. The photons are required to satisfy the following criteria: (1) the absolute value of the cosine of the polar angle (the angle between the photon and the beam directions) should be less than 0.95 to exclude the region of ``hot'' (noisy) crystals in the endcaps close to the beampipe, (2) the photon energy must lie in the interval 0.05 GeV $< E_{\gamma} <$ 0.43 GeV, (3) the angle to the closest projected charged track should be greater than $15^{\circ}$, (4) the shower should not be a fragment of a larger shower, and (5) the  pattern of energy deposition should be consistent with the single photon hypothesis. Photons satisfying these requirements are combined into pairs to form $\pi^0$ candidates. Combinations with momentum greater than 0.5 GeV/$c$ are excluded from further consideration. The pair of $\pi^0$'s remaining with the minimal value of the pull $\sqrt{S_{\gamma_{1}\gamma_2}^2+S_{\gamma_{3}\gamma_{4}}^2}$, where $S_{\gamma\gamma}=(m_{\gamma\gamma}-m_{\pi^0})/\sigma_{m_{\gamma\gamma}}$\ is then selected, and the missing mass calculated (Figure~\ref{figs:excl2pi0fit}). As is the case with charged pions we see clean signals in both lepton channels. Because of the poorer momentum resolution of reconstructed $\pi^0$'s than that of charged $\pi$'s, the distributions are considerably wider.

\begin{table}[hbt]
\center
\caption[]{\small Numbers of events observed after background subtraction, efficiencies, product of branching ratios ${\cal B}(\dectwo)\cdot{\cal B}(\deceigh)$ and branching ratio ${\cal B}(\dectwo)$ for the exclusive measurement.}\label{tab:excl2pi0}
\begin{tabular}{lcccc}
 & $N^{observed}$ & $\epsilon$ (\%) & ${\cal B}_{\pi\pi}\cdot {\cal B}_{ll}\ (\times 10^{-3})$ & ${\cal B}_{\pi\pi}$ \\ \hline
 $ee$ & $ 133.2\pm  11.5 $&$ 12.3\pm 1.0 $&$ 2.2\pm 0.2\pm 0.2 $&$ 0.088\pm 0.008\pm 0.010 $ \\
 $\mu\mu$ & $ 142.5\pm  11.9 $&$ 12.2\pm 1.0 $&$ 2.4\pm 0.2\pm 0.2 $&$ 0.096\pm 0.008\pm 0.009 $ \\
\end{tabular}
\end{table}

Once again we perform a side-band subtraction\footnote{Here the side-bands are $(9.10,9.40)\cup (9.55,9.80)$ GeV in both channels and the signal region is (9.40,9.55) GeV.} to extract the number of observed events (we estimate the background to be \Nbkgreltwo\  events, or \Nbkgrpreltwo\%, in the $ee$ channel and \Nbkgrmutwo\ events, or \Nbkgrprmutwo\%, in the $\mu\mu$ channel).

The yields and efficiencies for exclusive \dectwo,\deceigh\ transitions are presented in Table~\ref{tab:excl2pi0}. From these numbers we calculate the product of branching ratios ${\cal B}(\dectwo)\cdot {\cal B}(\deceigh)=N^{observed}_{ll}/(\epsilon_{ll}N_{\Utwo})$, and using the PDG values for \deceigh,\ we find ${\cal B}(\dectwo)$ which is also reported in Table~\ref{tab:excl2pi0}. Averaging over the two dilepton channels, we obtain:
\[ {\cal B}(\dectwo)=\BRtwo \]
In Table~\ref{tab:exclpi0comp} previous determinations of ${\cal B}(\dectwo)\cdot{\cal B}(\deceigh)$ are compared. From our two exclusive measurements we find the ratio ${\cal B}(\Utwo\rightarrow \Uone\pi^0\pi^0)/{\cal B}(\Utwo\rightarrow \Uone\pi^+\pi^-) =0.49\pm 0.06$ which is close to the isospin zero expectation of 0.53.

\begin{table}[hbt]
\center
\caption[]{\small ${\cal B}(\dectwo)\cdot{\cal B}(\deceigh)$ in units of $10^{-3}$.}\label{tab:exclpi0comp}
\begin{tabular}{ll}
ARGUS~\cite{ARGUS87}          & $2.3\pm 0.4\pm 0.5$ \\
Crystal Ball~\cite{CBALL85}   & $2.3\pm 0.3\pm 0.3$ \\
CUSB~\cite{CUSB84}            & $2.9\pm 0.5\pm 0.3$ \\
 this analysis & $ 2.29\pm 0.14\pm 0.20 $ \\ \hline
 average         & $ 2.34 \pm 0.19 $ \\
\end{tabular}
\end{table}

An inclusive analysis of the \dectwo\ transition gave a numerically consistent result, however because of the enormous combinatoric background, this measurement has very little statistical weight.

\section{Trigger efficiency and systematic errors}

The trigger system of the CLEO II detector, described in detail elsewhere~\cite{cleotrigger}, was designed for efficient triggering of two-photon, tau-pair, and hadronic events. There were eight active trigger lines in total during the \Utwo\ data taking, but only four of them are important in selecting events containing approximately back-to-back electron or muon pairs plus additional energy clusters in the calorimeter. To fire, these trigger lines require either two hits in the opposite hemispheres in the time-of-flight system or in the calorimeter, or a hit in the time-of-flight barrel region plus a track in the vertex detector (with small variations from line to line). Our estimates of the overall trigger efficiencies from a Monte Carlo simulation of the trigger system are reported in Table~\ref{tab:trigeff}.

\begin{table}[hb]
\center
\caption[]{\small Trigger efficiencies.}\label{tab:trigeff}
\begin{tabular}{lccccc}
        &\multicolumn{3}{c}{\decone}  &\multicolumn{2}{c}{\dectwo} \\ \cline{2-6}
  &$\Upsilon(1S)\rightarrow ee$ & $\Upsilon(1S)\rightarrow \mu\mu$ & $\Upsilon(1S)\rightarrow X$ & $\Upsilon(1S)\rightarrow ee$ & $\Upsilon(1S)\rightarrow \mu\mu$  \\ \hline
Efficiency & $0.961 \pm 0.008$ & $0.962\pm 0.015$& $0.990\pm 0.011$ & $0.982\pm 0.031$ & $0.977\pm 0.042$ \\
\end{tabular}
\end{table}

\begin{table}[hbt]
\center
\caption[]{\small Sources and magnitudes of systematic errors.}\label{tab:systerrors}
\begin{tabular}{lccc}
        &\multicolumn{3}{c}{Systematic error (\%)} \\ \cline{2-4}
        &\multicolumn{2}{c}{ \decone }  & \dectwo  \\ \cline{2-4}
Source  & Exclusive & Inclusive & Exclusive \\ \hline
Multiplicity of event     & --- & 2.0  & ---   \\ 
Trigger efficiency        & 0.9 / 1.6$^{a}$ & 1.1  & 3.1 / 4.2 \\
Tracking                  & 2.8 & 2.8  & ---   \\ 
$\pi^0$-finding         & --- & ---  &  7.0  \\ 
Finite MC sample          & 0.5 & 0.5 & 0.5  \\ 
Background subtraction    & \dNbkgrelone\ / \dNbkgrmuone\ & ---  & \dNbkgreltwo\ / \dNbkgrmutwo\  \\
Leptonic branching ratios & 6.7 / 2.8 & --- & 6.7 / 2.8 \\
Fitting function          & --- & 0.5  &  ---  \\ 
$N_{\Utwo}^{prod}$        & 3.7 & 3.7 & 3.7  \\ \hline
Total                     & 8.2 / 5.7 & 5.2  &  10.9 / 9.4 \\
\end{tabular}
\raggedright {\small $^{a}$ separately for $ee/\mu\mu$ channels.}
\end{table}

The dominant systematic errors in our analysis come from uncertainties in the total number of produced \Utwo\ resonance events, the leptonic branching fractions of the \Uone, and the charged track and $\pi^0$ finding efficiency. Other systematic errors are due to uncertainties in trigger efficiencies, event environment effects, background subtraction, and the shape of the fitting function (inclusive analysis only). The complete breakdown of systematic errors is given in Table~\ref{tab:systerrors} (relative errors in percent). All these errors are considered to be uncorrelated and separately contribute to systematic uncertainties in our branching ratios. 

\section{Dipion invariant mass spectra in \boldmath{$\Utwo\rightarrow \Uone\pi\pi$} transitions}

\begin{figure}[htb]
\center
\epsfig{file=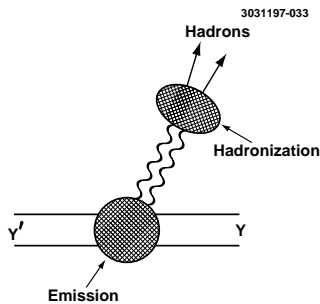, height=4cm}
\caption[]{A hadronic transition as a two-step process.}\label{figs:hadrdiag}
\end{figure}

There have been several theoretical predictions for the dipion invariant mass distribution since a significant difference from phase space was found in $\psi'\rightarrow J/\psi\pi\pi$ transitions~\cite{psispectrum}. As shown in Figure~\ref{figs:hadrdiag}, the dipion transition is treated as a factorizable two-step process: emission of gluons from heavy quarks and conversion of the gluons into light hadrons. The dipion invariant mass spectrum is determined by the second step, in the hadronization of two gluons emitted by the decaying bottomonium -- a process which is not well understood.

The following parameterizations were used in fitting our experimental distributions:
\begin{itemize}
\item {Yan~\cite{Yanmodified} model: 
\[ \frac{d\sigma}{dm_{\pi\pi}}\propto PS\cdot [(m_{\pi\pi}^2-2m_{\pi}^2)^2+\frac{B}{3A}(m_{\pi\pi}^2-2m_{\pi}^2)\cdot (m_{\pi\pi}^2-4m_{\pi}^2+2K^2(1+\frac{2m_{\pi}^2}{m_{\pi\pi}^2}))+O(\frac{B^2}{A^2})]\] 
where $K=\frac{M_2^2-M_{1}^2+m_{\pi\pi}^2}{2M_2}$ } 
\item {Voloshin and Zakharov~\cite{Vol-Zakh} model: 
\[ \frac{d\sigma}{dm_{\pi\pi}}\propto PS\cdot[m_{\pi\pi}^2-\lambda m_{\pi}^2]^2\] }
\item {Novikov and Shifman~\cite{Nov-Shif} model: 
\[ \frac{d\sigma}{dm_{\pi\pi}}\propto PS\cdot [m_{\pi\pi}^2-k(M_2-M_{1})^2(1-\frac{2m_{\pi}^2}{m_{\pi\pi}^2})+O(k^2)]^2\] }
\end{itemize}
In all the above formulas $M_2=M_{\Utwo},\ M_{1}=M_{\Uone}$ and $PS$ is the phase space factor:
\[ PS=\sqrt{ \frac{ (m_{\pi\pi}^2-4m_{\pi}^2) [M_{1}^{4}+M_2^{4}+m_{\pi\pi}^{4}-2(M_{1}^2m_{\pi\pi}^2+M_2^2m_{\pi\pi}^2+M_{1}^2M_2^2) ] } {4M_2^2} } \]

\subsection{\em The $\pi^+\pi^-$ invariant mass spectrum}

\begin{figure}[htb]
\center
\epsfig{file=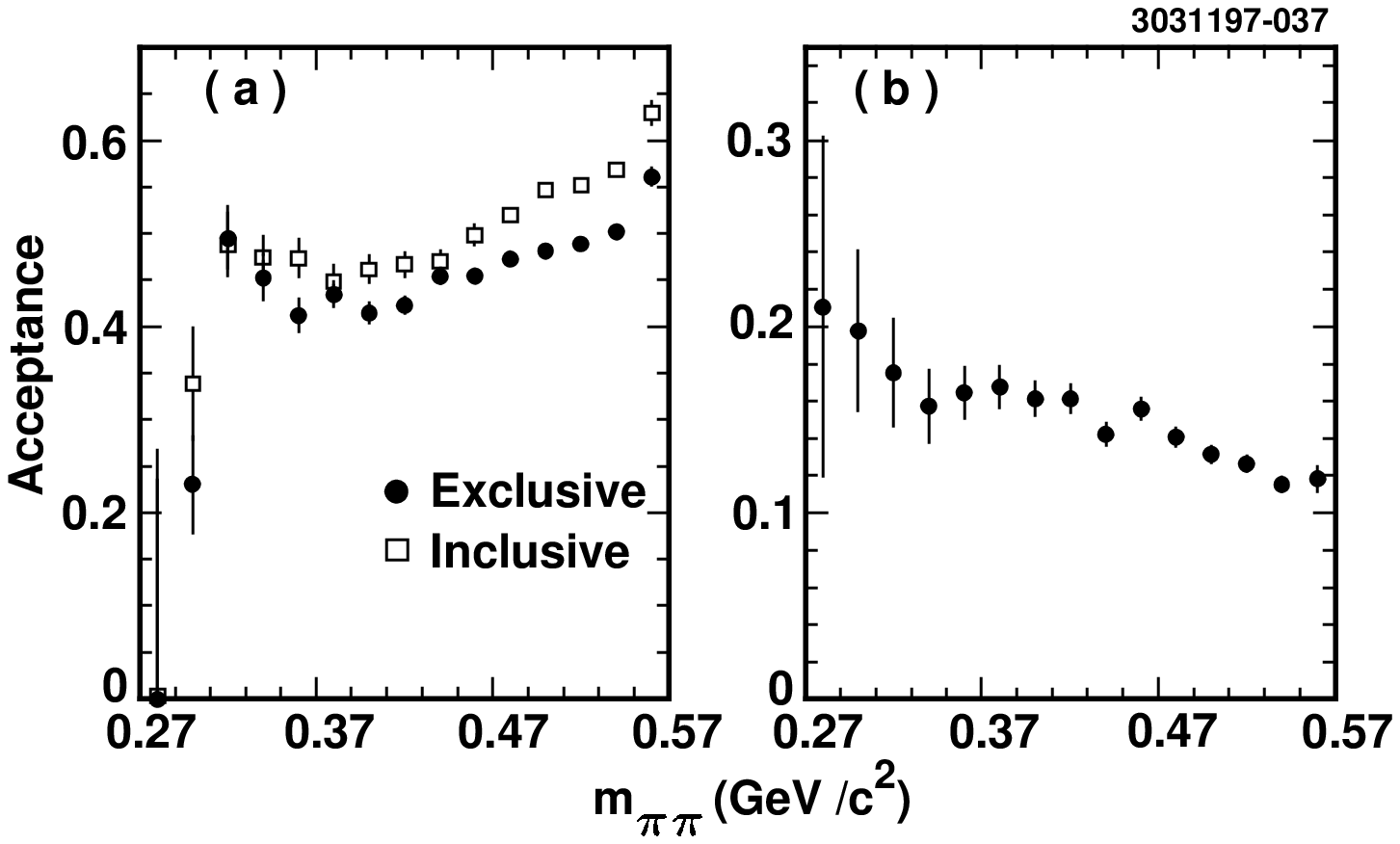, height=8cm}
\caption[]{Dipion invariant mass acceptance for a) $\Utwo\rightarrow \Uone\pi^+\pi^-$ and b) $\Utwo\rightarrow \Uone\pi^0\pi^0$ events.}\label{figs:invmsaccept}
\end{figure}

\begin{figure}[htb]
\center
\epsfig{file=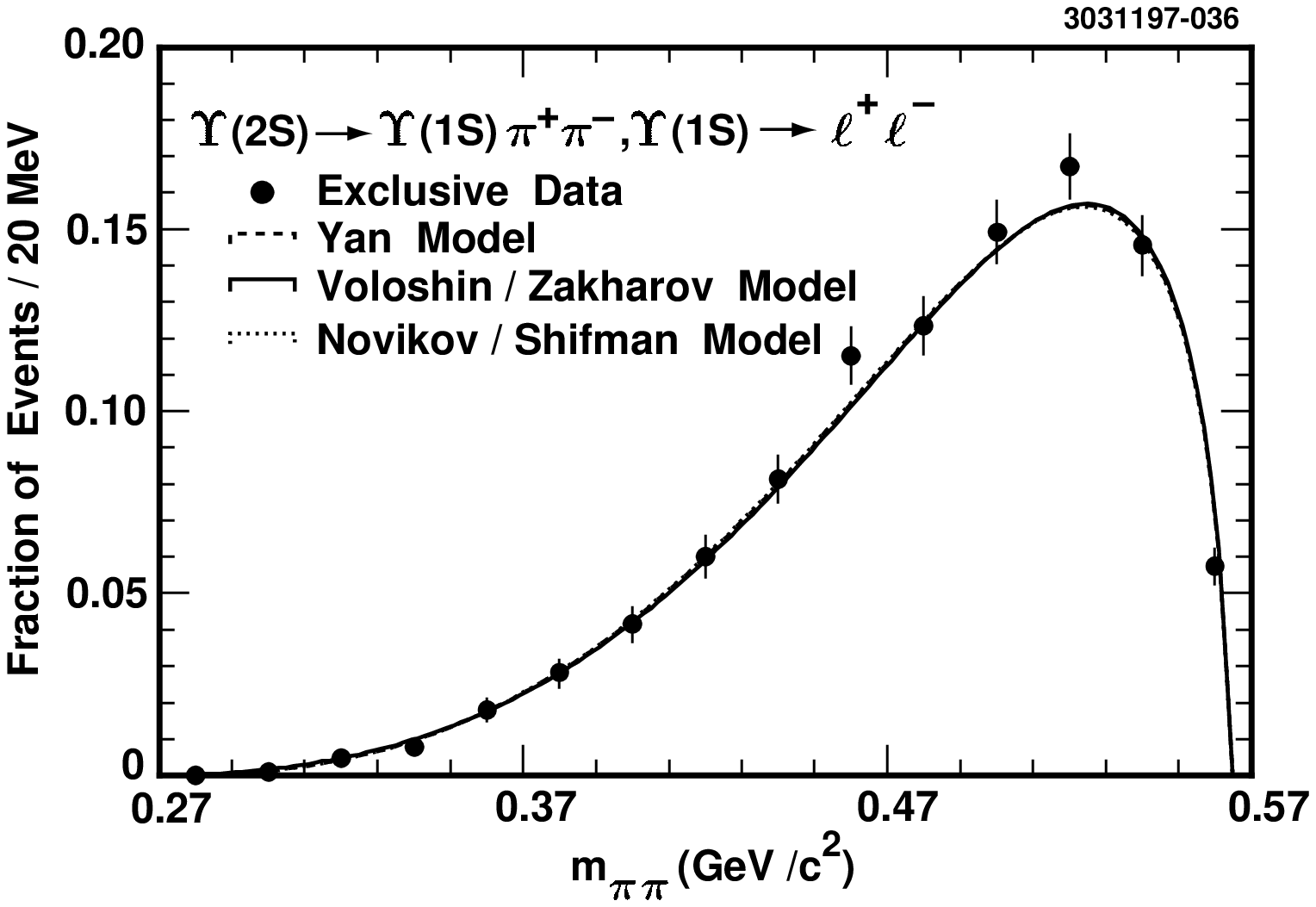, height=9cm}
\caption[]{Dipion invariant mass spectrum from exclusive \decone\ events\\ (corrected for acceptance).}\label{figs:excl2piinvms}
\end{figure}

We extract a dipion invariant mass spectrum from both the inclusive and exclusive samples. The dipion invariant mass spectrum from exclusive events is shown in Figure~\ref{figs:excl2piinvms}, where we have combined results from both $ee$ and $\mu\mu$ channels. The data points in this histogram are the sideband-subtracted yields for the corresponding bins in $m_{\pi\pi}$, where each data point has been corrected for acceptance (Figure~\ref{figs:invmsaccept}). The fits to this spectrum, using the aforementioned parameterizations are also shown in Figure~\ref{figs:excl2piinvms}; they are all consistent with our data. 

\begin{figure}[htb]
\center
\epsfig{file=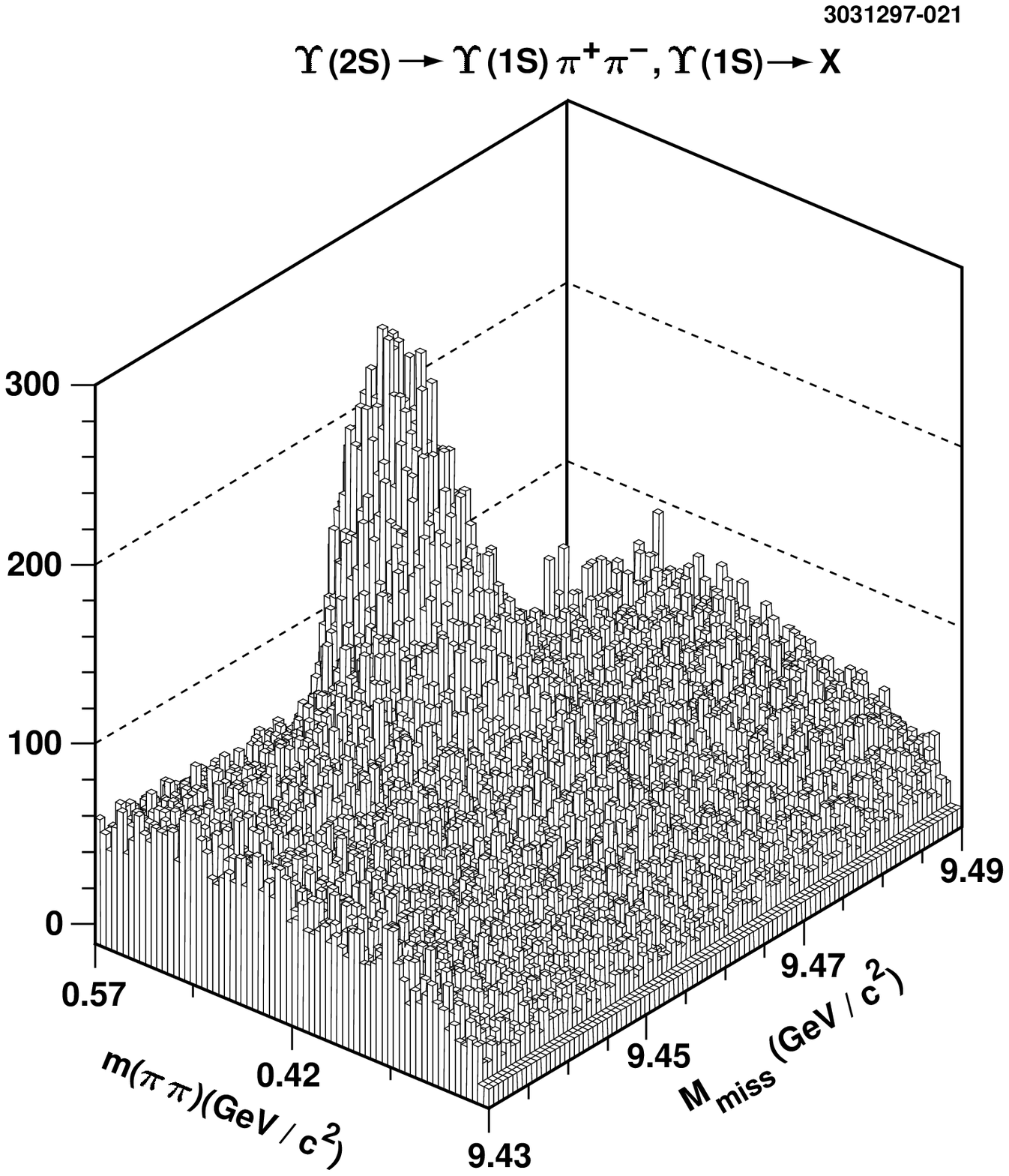, height=10cm}
\caption[]{Dipion invariant mass vs. missing mass from the inclusive $\pi^+\pi^-$ events.}\label{figs:inclm2pivsmm}
\end{figure}

\begin{figure}[htb]
\center
\epsfig{file=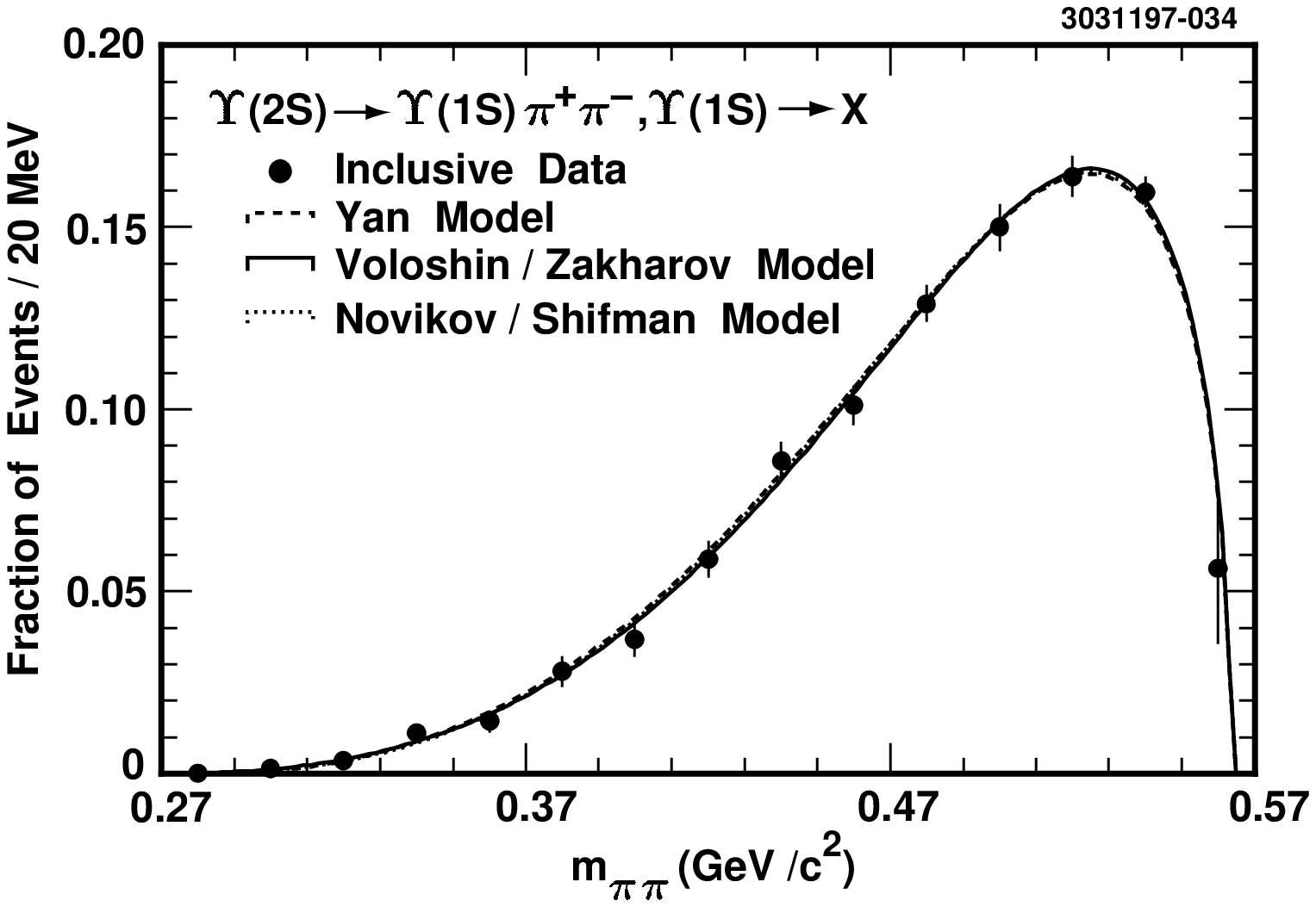, height=9cm}
\caption[]{Dipion invariant mass spectrum from inclusive \decone\ events\\ (corrected for acceptance).}\label{figs:incl2piinvms}
\end{figure}

To produce the dipion invariant mass spectrum in the inclusive measurement, we use a two-dimensional plot of $m_{\pi\pi}$ vs. $M_{miss}$ (shown in Figure~\ref{figs:inclm2pivsmm}) which we slice in bins of $m_{\pi\pi}$, project onto the $M_{miss}$ axis and then fit the projections with a double Gaussian for the \Uone\ peak plus a third order polynomial to represent the background. We correct the fitted number of \Uone\ events for acceptance on a bin-by-bin basis (Figure~\ref{figs:invmsaccept}) to obtain the dipion invariant mass spectrum. Fits to the the different parameterizations of this distribution are given in Figure~\ref{figs:incl2piinvms}. In Table~\ref{tab:dpms2pipar} we have compiled the values of the fitting parameters, their errors, and $\chi^2$ values of the fits for both the exclusive and inclusive measurements.

\begin{table}[h]
\center
\caption[]{\small Fit results for the $\pi^+\pi^-$ invariant mass spectrum parameterizations.}\label{tab:dpms2pipar}
\begin{tabular}{lcccc}
    & \multicolumn{2}{c}{Exclusive events} &\multicolumn{2}{c}{Inclusive events} \\ \cline{2-5}                 
Model & Fit parameters & $\chi^2$/\invmassdf d.f. & Fit parameters & $\chi^2$/\invmassdf d.f.\\ \hline
 Yan~\cite{Yan} & $B/A=-0.132\pm  0.018$ &  15.6 & $-0.154\pm  0.014$ &   9.3\\
 Voloshin and Zakharov~\cite{Vol-Zakh} & $\lambda=  3.11 \pm  0.18 $ &  17.5 & $ 3.42\pm  0.16$ &   6.6\\
 Novikov and Shifman~\cite{Nov-Shif} & $k= 0.138\pm  0.009$ &  15.1 & $ 0.153\pm  0.008$ &   8.4\\
\end{tabular}
\end{table}

\subsection{\em The $\pi^0\pi^0$ invariant mass spectrum}

\begin{figure}[htb]
\center
\epsfig{file=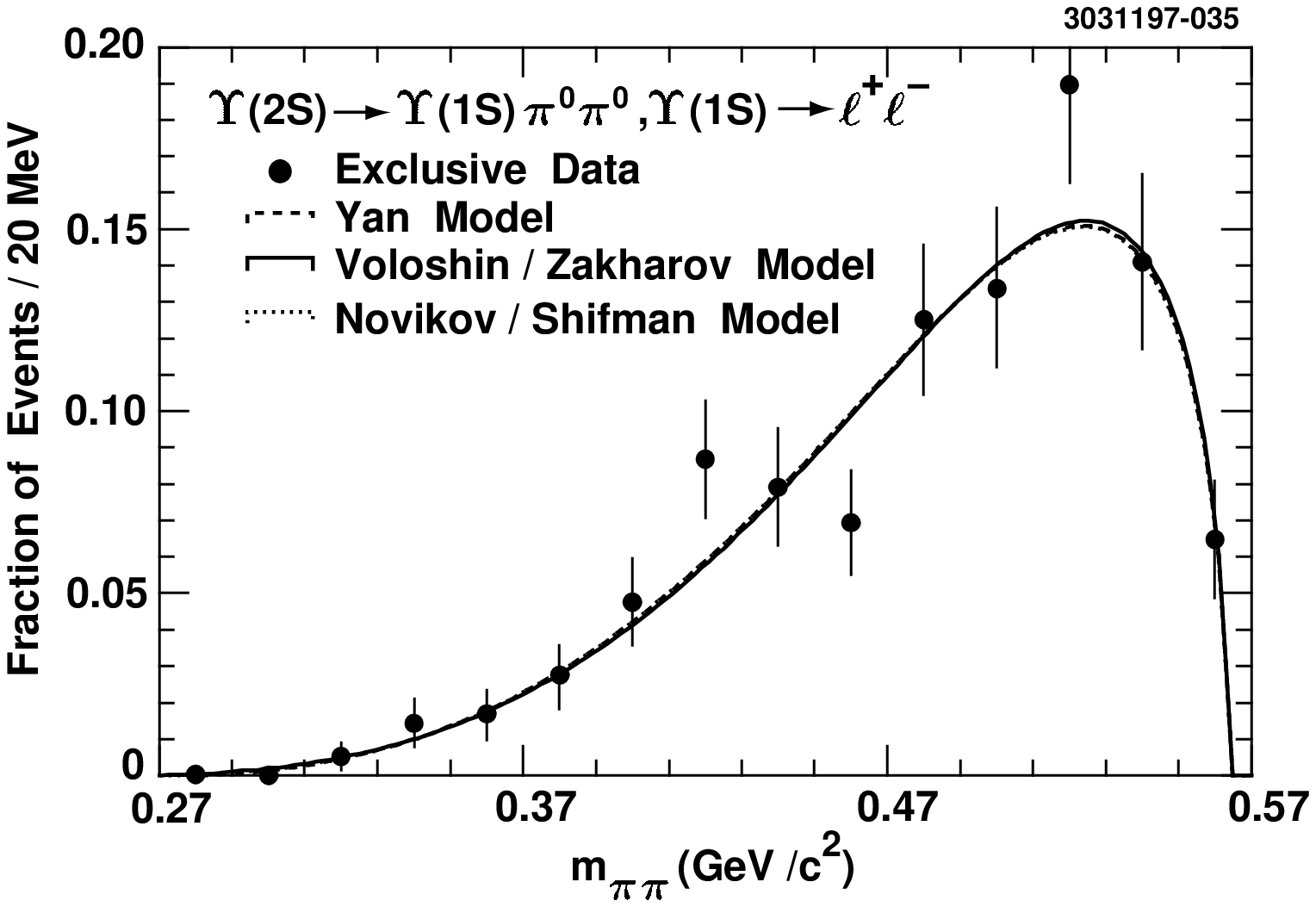, height=9cm}
\caption[]{Dipion invariant mass spectrum from exclusive \dectwo\ events\\ (corrected for acceptance).}\label{figs:excl2pi0invms}
\end{figure}

We have a sample of \dectwo\ exclusive events, large enough to produce the $\pi^0\pi^0$ invariant mass spectrum for this transition. Here too the data points in our histogram are the yields of exclusive $l^+l^-\pi^0\pi^0$ events in each corresponding $m_{\pi\pi}$ bin, corrected for acceptance 
\begin{table}[hb]
\center
\caption[]{\small Fit results for the $\pi^0\pi^0$ invariant mass spectrum parameterizations.}\label{tab:dpms2pi0par}
\begin{tabular}{lcc} 
        & \multicolumn{2}{c}{Exclusive events} \\ \cline{2-3}
 Model & Fit parameters & $\chi^2$/\invmassdf d.f.\\ \hline
 Yan~\cite{Yan} & $B/A=-0.145\pm  0.040$ & 10.8\\
 Voloshin and Zakharov~\cite{Vol-Zakh} & $\lambda=  3.35 \pm  0.49 $ & 11.1\\
 Novikov and Shifman~\cite{Nov-Shif} & $k= 0.139\pm  0.022$ & 10.9 \\
\end{tabular}
\end{table}
(see Figure~\ref{figs:invmsaccept}). The fits to the acceptance-corrected $\pi^0\pi^0$ invariant mass spectrum are shown in Figure~\ref{figs:excl2pi0invms}, with fit results reported in Table~\ref{tab:dpms2pi0par}.

\subsection{\em Combined results for the $\pi\pi$ invariant mass measurements}

\begin{figure}[htb]
\center
\epsfig{file=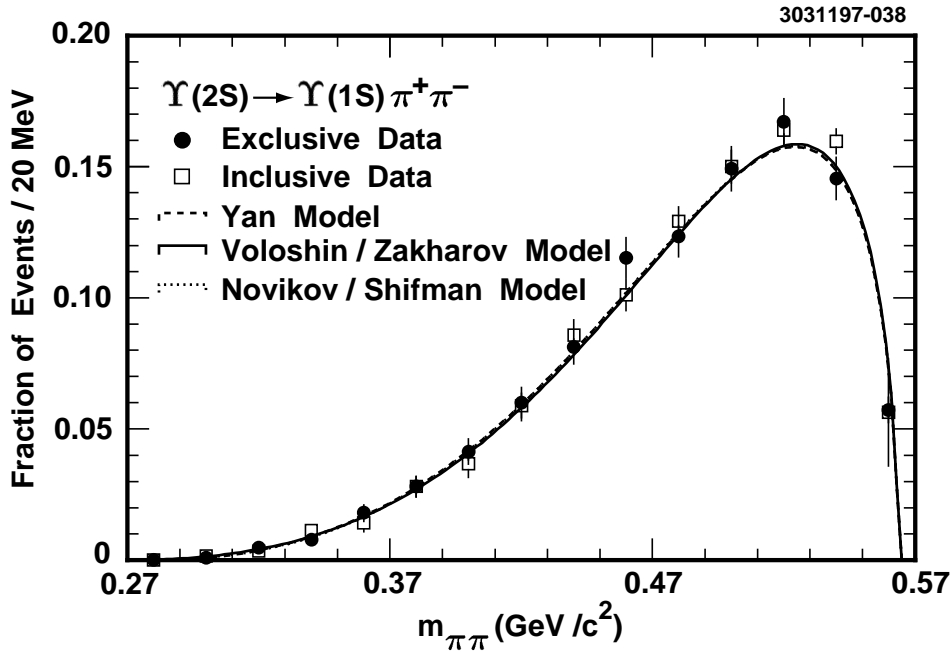, height=9cm}
\caption[]{Combined fit to the dipion invariant mass spectrum from exclusive and inclusive \decone\ events.}\label{figs:combinvms}
\end{figure}

In order to compare the results of our analysis with the results of other experiments, we perform a simultaneous fit to the exclusive and inclusive $\pi^+\pi^-$ invariant mass spectra. We do not include the $\pi^0\pi^0$ measurement in the combined fit because it has a slightly different parameterization (due to the difference in mass between neutral and charged pions) and much lower statistical significance. The fits to the combined data of the exclusive and inclusive \decone\ decays are shown in Figure~\ref{figs:combinvms}. In Table~\ref{tab:dpmsparavrg} we compare the results of our combined fit with the results from previous experiments. 

\begin{table}[h]
\center
\caption[]{\small Values of fit parameters using different parameterizations of the $\pi\pi$ invariant mass spectrum.}\label{tab:dpmsparavrg}
\begin{tabular}{lccc}
Model & Yan~\cite{Yan} & Voloshin and Zakharov~\cite{Vol-Zakh} & Novikov and Shifman~\cite{Nov-Shif} \\ \hline
Parameter & $B/A$ & $\lambda$ & $k$ \\ \hline
Crystal Ball~\cite{CBALL85} & $-0.18\pm 0.15$ & $3.3\pm 1.2$ & $0.14\pm 0.05$ \\
CLEO~\cite{CLEO84} & $-0.18\pm 0.06$ & $3.2\pm 0.4$ & $0.15\pm 0.02$ \\
ARGUS~\cite{ARGUS87}& $-0.154\pm 0.019$ & $3.30\pm 0.19$ & $0.151\pm 0.009$ \\
 this analysis & $-0.145\pm  0.011$ & $  3.28 \pm  0.12$ & $ 0.146\pm  0.006$ \\
\end{tabular}
\end{table}

\section{Angular distributions}

\begin{figure}[htb]
\center
\epsfig{file=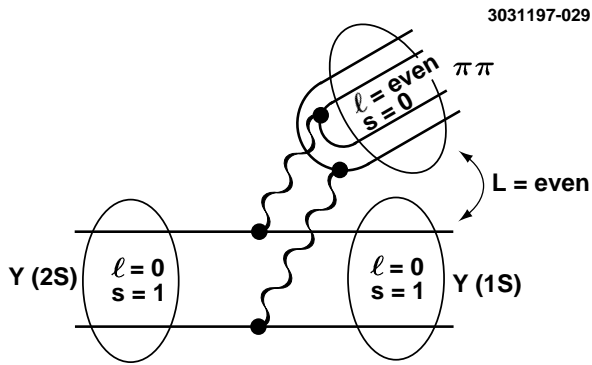, height=5cm}
\caption[]{Angular momenta in the $\pi\pi$ transitions.}\label{figs:angqnum}
\end{figure}

The angular distributions in $\pi\pi$ transitions were studied using our exclusive and inclusive $\pi^+\pi^-$ data samples. In $e^+e^-$ annihilation the \Utwo\ is produced polarized with its spin axis lying along the beam axis. This total angular momentum (and its projection onto the beam axis) must be conserved. There are three possible angular momenta in the final state of the dipion transition (Figure~\ref{figs:angqnum}): the total spin of the \Uone\ ${\bf J}$, the internal orbital angular momentum of the dipion system ${\bf l}$ (the total spin of the dipion system $s=0$) and the orbital angular momentum of the dipion system relative to the \Uone\ ${\bf L}$~\cite{Cahn}.

\begin{figure}[hbt]
\center
\epsfig{file=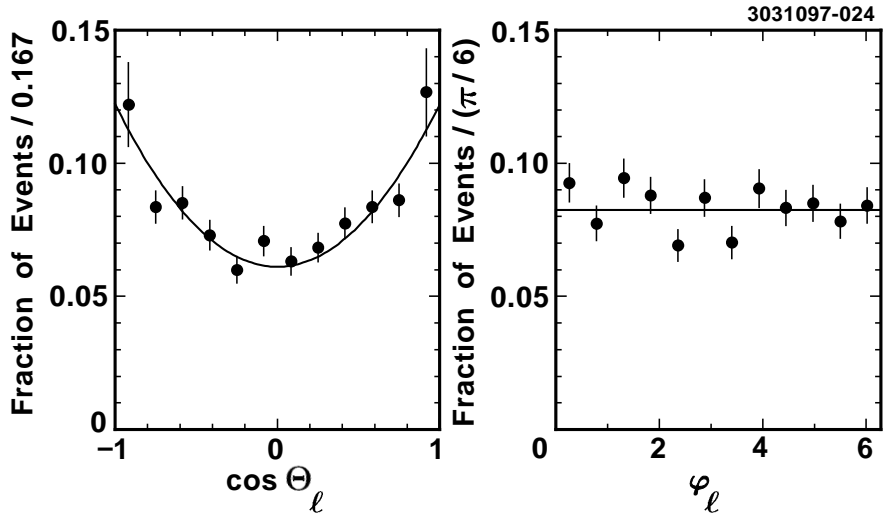, height=7cm}
\caption[]{Angular distributions of $l^+$ from \deceigh\ in the center of mass frame (corrected for acceptance). Solid lines are $dN/d(\cos \theta_{l})=N\cdot(1+\cos^2\theta_{l})$ and $dN/d\phi_{l}=const$ fits.}\label{figs:angdislep}
\end{figure}

Since the transition is expected to be dominated by $E1\cdot E1$ gluon radiation, the angular momentum of the $b\overline{b}$ system is not changed by the dipion decay and the polarization of the parent \Utwo\ should be observed in the subsequent decay of the daughter \Uone. This is verified in the $\cos\theta$ and $\phi$ distributions of the outgoing $l^+$ with respect to the beam shown in Figure~\ref{figs:angdislep}: the expected $(1+\cos^2\theta)$ distribution is clearly seen in the $\cos\theta_{l^+}$ distribution and the azimuthal distribution $\phi_{l^+}$ is reasonably flat.\footnote{It should be flat because CESR beams do not have azimuthal polarization.}

\begin{figure}[htb]
\center
\epsfig{file=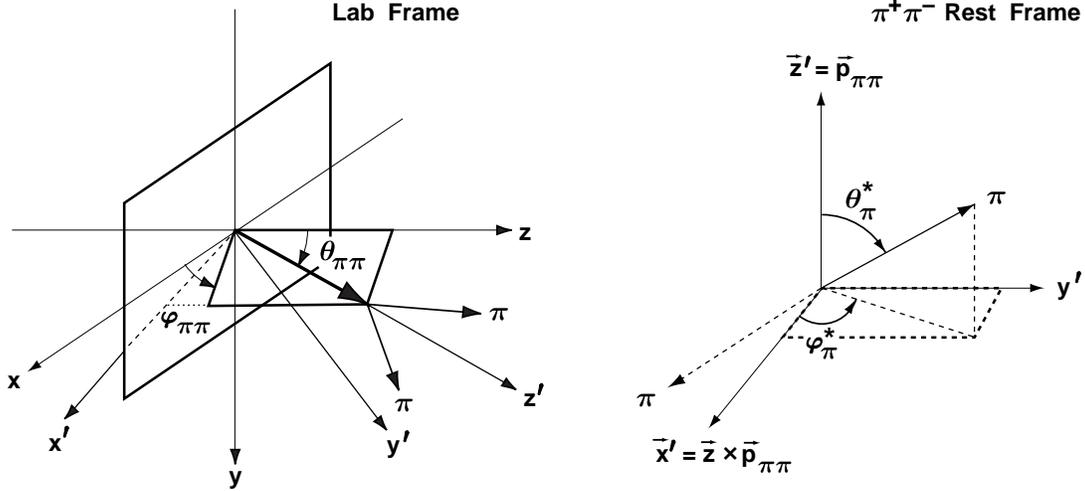, height=7cm}
\caption[]{Frames of reference and definitions of angles for the $\pi\pi$ transitions.}\label{figs:angframes}
\end{figure}

The quantum numbers of both the \Utwo\ and \Uone\ are $J^{PC}=1^{--}$ and $I^{G}=0^-$; the dipion system has $I^{GC}=0^{++}$. Parity forces $l$ and $L$ to be both even or both odd. The $G$-parity for the dipion system\footnote{The operation of charge conjugation followed by isospin rotation does not change the state of the dipion system.} is 1 and from the formula $G=(-1)^{l+s+I}$ with $I=0,\ s=0,\ G=1$ we find that $l$, hence $L$, must be even.

\begin{figure}[htb]
\center
\epsfig{file=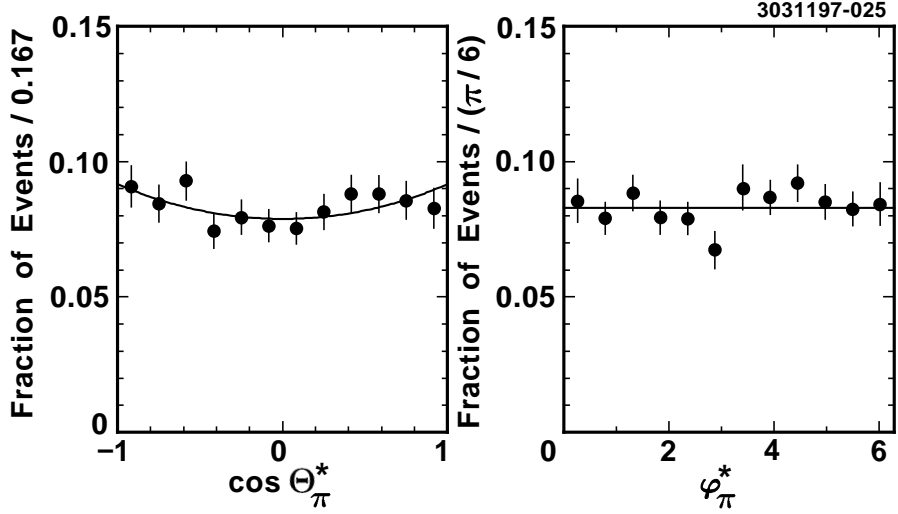, height=7cm}
\caption[]{ $\cos\theta^{*}_{\pi}$ and $\phi^{*}_{\pi}$ distributions of $\pi^+$ in the center of mass frame of $\pi^+\pi^-$ system in the exclusive \decone\ measurement (corrected for acceptance). Solid lines are $dN/d(\cos\theta^{*}_{\pi})=N\cdot|\sqrt{1-\epsilon^2}Y^{0}_{0}+\epsilon Y^{0}_{2}| ^2$ and $dN/d\phi^{*}_{\pi}=const$ fits.}\label{figs:angdispihel}
\end{figure}

\begin{figure}[hbt]
\center
\epsfig{file=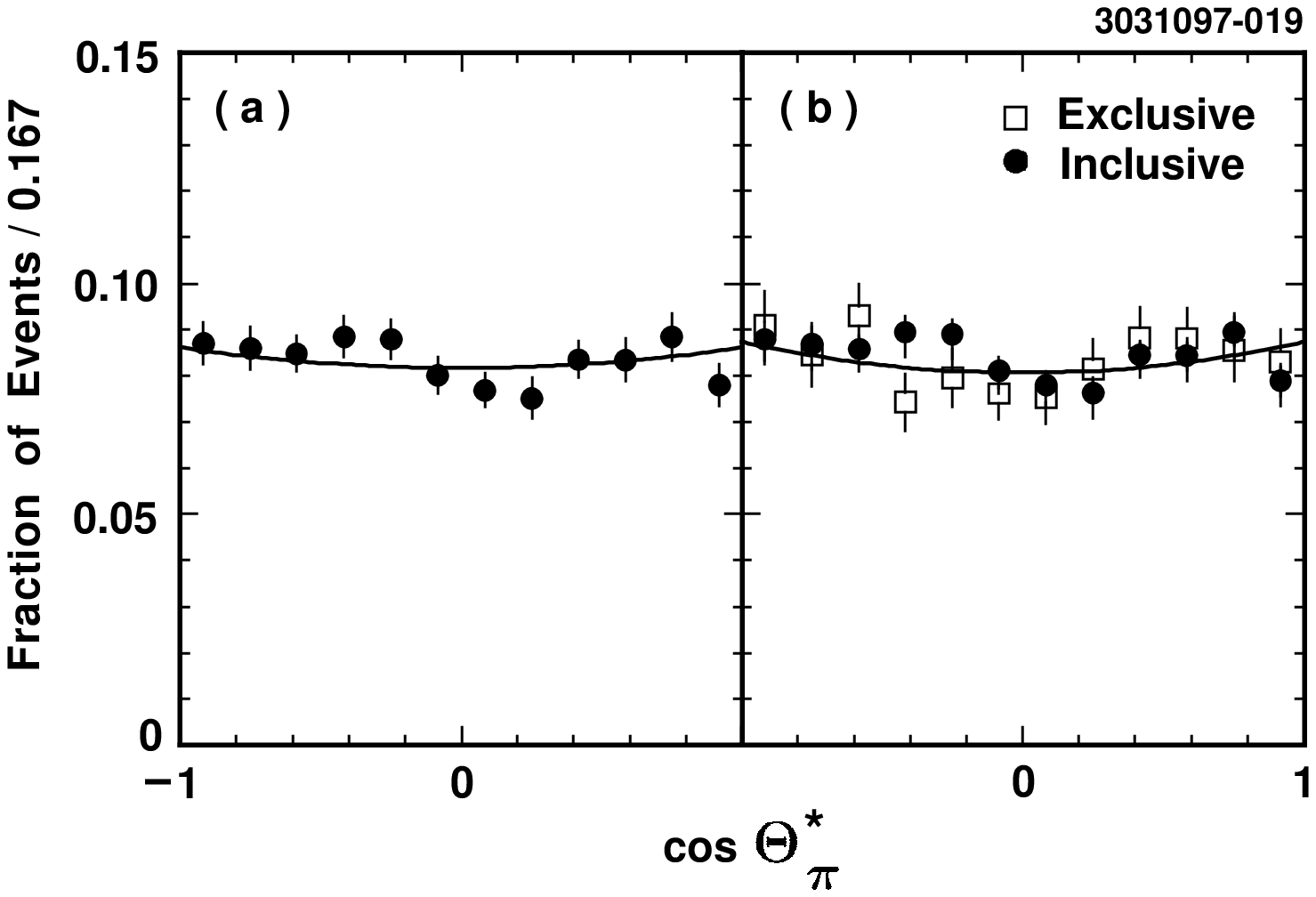, height=7cm}
\caption[]{a) Fit to the $\cos\theta^{*}_{\pi}$ distribution in the inclusive \decone\ measurement. b) Combined fit to the $\cos\theta^{*}_{\pi}$ distributions (corrected for acceptance).}\label{figs:2pi-angdispihel-comb}
\end{figure}

All theoretical models describing the dipion invariant mass spectrum predict the pions to be emitted predominantly in the $s$-wave state ($l=0$), although there exists a prediction for the $d-$wave contribution ($l=2$)~\cite{Nov-Shif} of the order of 1\%. The $d-$wave contribution can be observed in the $\cos\theta^{*}_{\pi}$ distribution of the angle $\theta^{*}_{\pi}$ of the $\pi^+$ in the $\pi\pi$ center of mass frame with respect to the $\pi\pi$ direction. (See Figure~\ref{figs:angframes} for definitions of angles.) This is shown in Figure~\ref{figs:angdispihel} along with the $\phi^{*}_{\pi}$ distribution which should be flat. It is possible to fit the $\cos\theta^{*}_{\pi}$ distribution for our exclusive data sample to a coherent sum of $s-$ and $d-$waves; $\epsilon$ here represents the size of the $d-$wave contribution:
\[\frac{dN}{d(\cos\theta^{*}_{\pi})}\propto | \sqrt{1-\epsilon^2}\; Y^{0}_{0}+\epsilon Y^{0}_{2}| ^2 \]
with the fit result: $\epsilon = \Epstwopitwol$. We studied whether this effect is present in our inclusive \decone\ sample (Figure~\ref{figs:2pi-angdispihel-comb}a). In the inclusive measurement the fit result is: $\epsilon = \Epstwopi$. Performing a simultaneous fit to the combined data from the exclusive and inclusive measurements (Figure~\ref{figs:2pi-angdispihel-comb}b) we find:
\[ \epsilon = \Epstwopicomb\]
Our results demonstrate the strong $s-$wave dominance expected in the dipion transition and show some indication of a $d-$wave contribution on the order of a few percent. In a similar analysis, ARGUS~\cite{ARGUS87} obtained $\epsilon = 0.018^{+0.108}_{-0.009}$.

\begin{figure}[ht]
\center
\epsfig{file=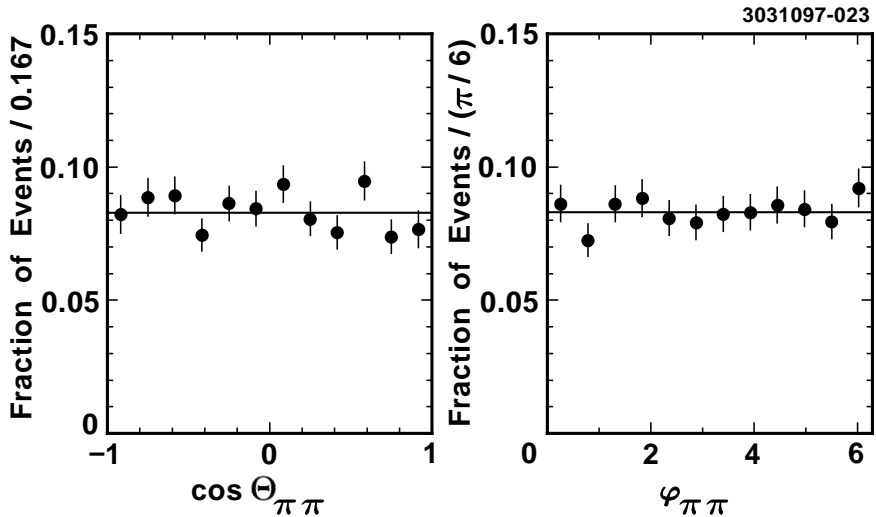, height=7cm}
\caption[]{$\cos\theta$ and $\phi$ distributions of $\pi^+\pi^-$ system in the $e^+e^-$ frame in the exclusive \decone\ measurement (corrected for acceptance).}\label{figs:angdis2pi}
\end{figure}

To examine further the question of a possible $d-$wave contribution we performed a fit to the combined data with the value of $\epsilon$\ fixed at zero and found the fit confidence level to be $\CLEpszero\%$. Using the values of the $\chi^2$ from the two combined fits, we performed the $F-$test\footnote{See, for example, Bevington, P.R. ``Data reduction and error analysis for the physical sciences.''} for the significance of the $d-$wave contribution. We calculate $F_{\chi}=\Delta \chi^2/\chi^2_{n}=3.5/0.929=3.77$ for $n=21$ d.f. which means that adding the $d-$wave to the fitting function significantly improves the fit and corresponds to a 7\% probability that the parent distribution does not have the $d-$wave term.

The spatial orientation of the $\pi\pi$ system in the \Utwo\ frame is consistent with isotropy (Figure~\ref{figs:angdis2pi}) which implies that there is no significant contribution from a ``relative'' $D-$wave ($L=2$). 

\section{Transition \boldmath{\decthree}}

In our analysis of this transition we used the decay modes where the \Uone\ decays into a lepton pair ($e$ or $\mu$) and the $\eta$ decays via one of the modes: $\eta\rightarrow 3\pi^0\rightarrow 6\gamma$, $\eta\rightarrow 2\gamma$, $\eta\rightarrow \pi^+\pi^-\pi^0 \rightarrow \pi^+\pi^-\gamma\gamma$, or $\eta\rightarrow  \pi^+\pi^-\gamma$ (the total branching ratio of these four modes is 98.2\%). The selection criteria common to the four $\eta$ decay modes are: (1) requirements on the leptonic pair which are the same as in our exclusive \decpiall\ analyses, (2) a requirement on the $\eta$ candidate momentum $p_{\eta} < 0.2$ GeV/$c$, and (3) a requirement on the dilepton invariant mass 9.21 GeV/$c^2 < m_{ll} < 9.71$ GeV/$c^2$.

\begin{figure}[tb]
\center
\epsfig{file=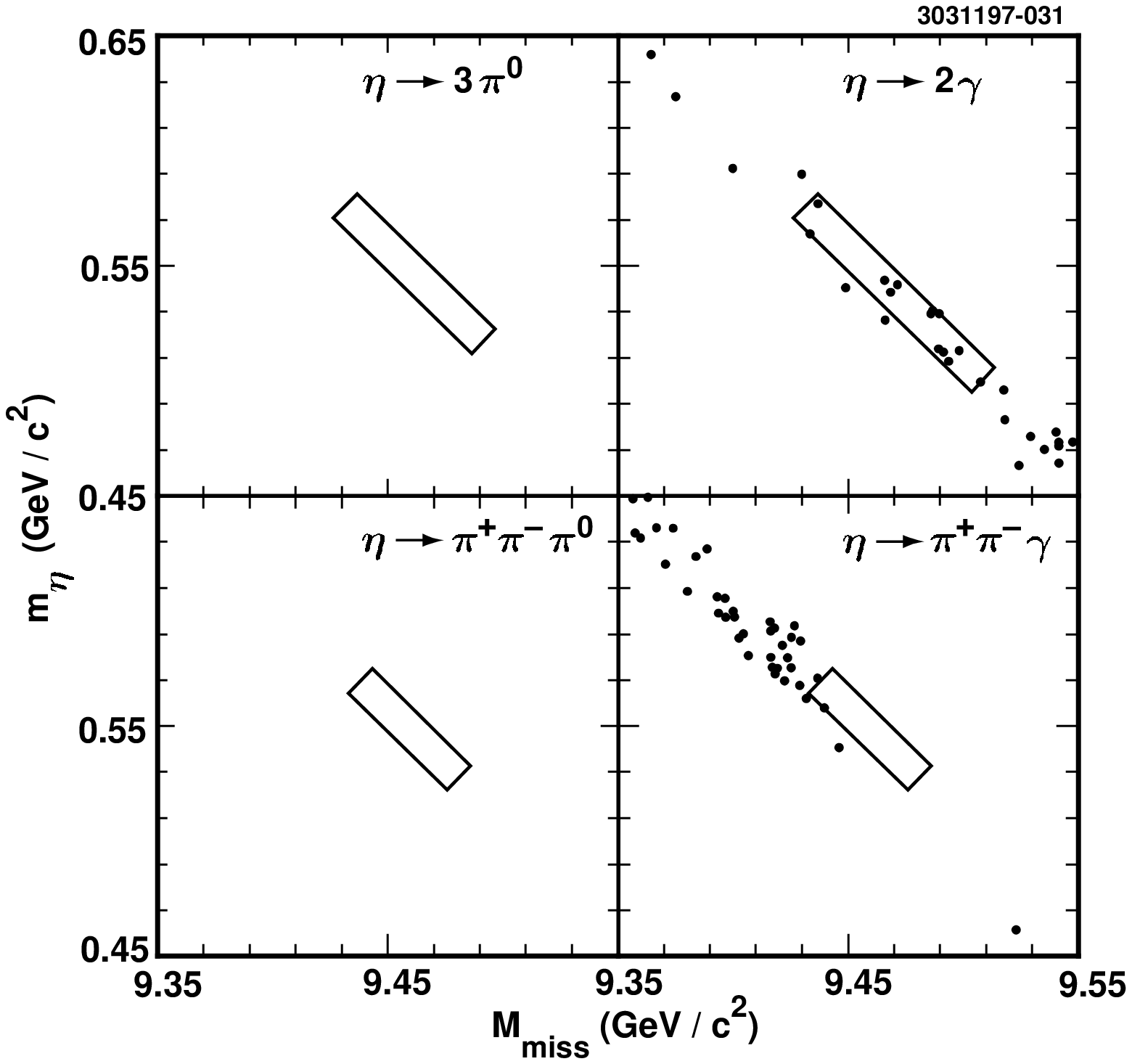, height=9.5cm}
\caption[]{Signal from \decthree,$\Uone\rightarrow e^+e^-$\ in different $\eta$ decay modes.}\label{figs:etascatdtel}
\end{figure}
\begin{figure}[hbt]
\center
\epsfig{file=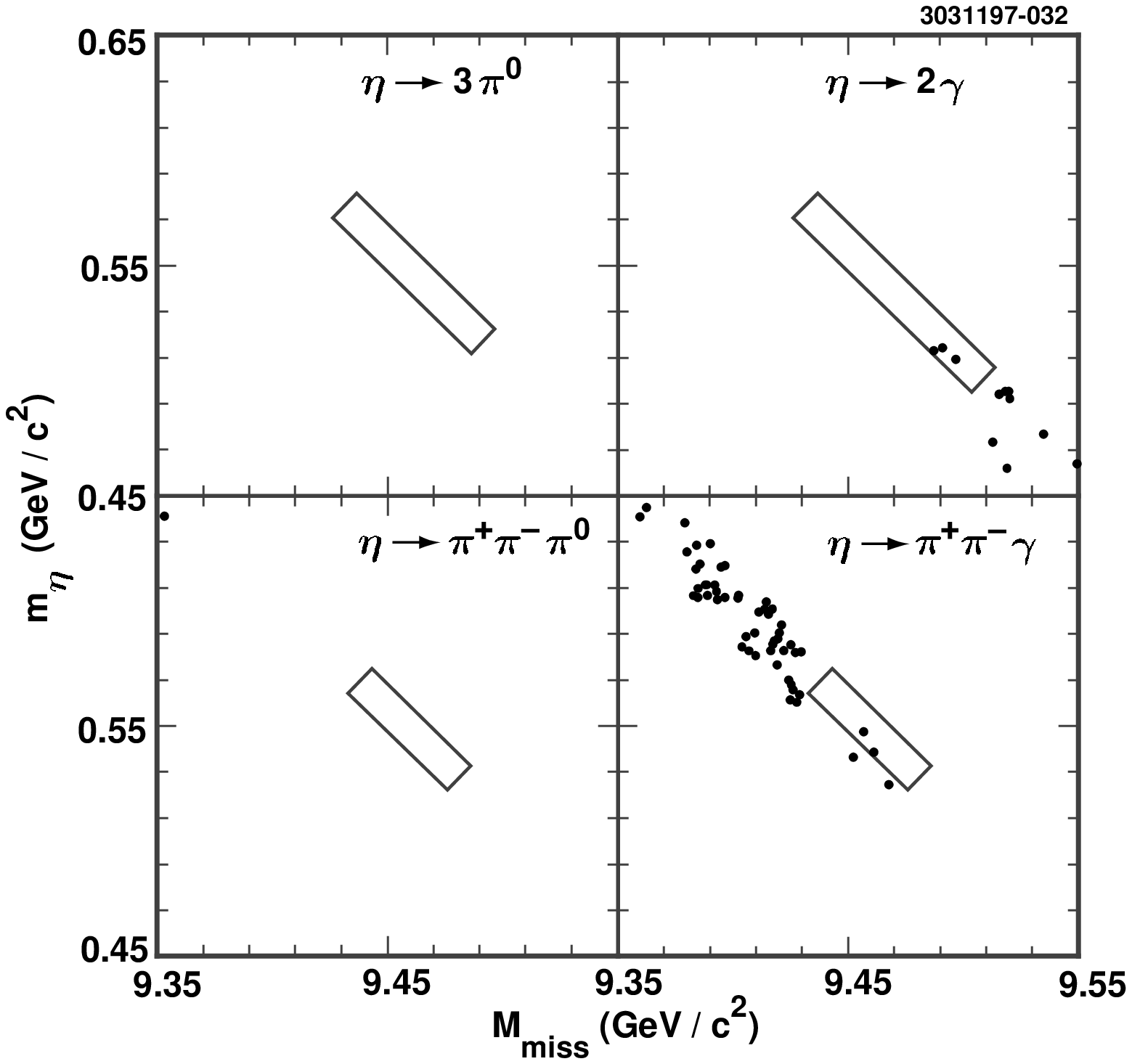, height=9.5cm}
\caption[]{Signal from \decthree,$\Uone\rightarrow \mu^+\mu^-$\ in different $\eta$ decay modes.}\label{figs:etascatdtmu}
\end{figure}

For the modes  $\eta\rightarrow 3\pi^0\rightarrow 6\gamma$ and $\eta\rightarrow 2\gamma$ the following additional criteria are applied: (1) photon requirements as in the exclusive \dectwo\ analysis except that the energy of $\gamma$'s from $\eta\rightarrow 3\pi^0$ must satisfy  $E_{\gamma} < 0.33$ GeV  and those from $\eta\rightarrow 2\gamma$ must satisfy $E_{\gamma} < 0.6$ GeV, (2) there should be two good charged tracks in the event, (3) the number of showers in the calorimeter unmatched to charged tracks should be less then seven (for $\eta\rightarrow 3\pi^0$) or three (for $\eta\rightarrow 2\gamma$), (4) for $\eta\rightarrow 3\pi^0$ the $\pi^0$ candidate momentum must satisfy $p_{\pi^0} < 0.3$ GeV/$c$, and (5) for $\eta\rightarrow 2\gamma$ the cosine of the angle between the two photons must satisfy $cos\theta_{\gamma\gamma} < -0.85$ to reduce the background from the QED process $e^+e^-\rightarrow \gamma\gamma e^+e^-$ (since the $\eta$'s are produced almost at rest, the daughter $\gamma$'s are close to being back to back).

In the modes $\eta\rightarrow \pi^+\pi^-\pi^0 \rightarrow \pi^+\pi^-\gamma\gamma$ and $\eta\rightarrow  \pi^+\pi^-\gamma$ we require: (1) the charged pions must pass the same cuts as in the exclusive \decone\ measurement, (2) there must be four good charged tracks in the event, (3) there must be three (for $\eta\rightarrow \pi^+\pi^-\pi^0 \rightarrow \pi^+\pi^-\gamma\gamma$) or two (for $\eta\rightarrow  \pi^+\pi^-\gamma$) showers in the calorimeter unmatched to charged tracks, and (4) the cosine of the opening angle between charged pions must satisfy $cos\theta_{\pi^+\pi^-} < 0.9$ to suppress background from QED processes with gamma conversion $\gamma\rightarrow e^+e^-$ where the $e^+e^-$-pair fakes a $\pi^+\pi^-$-pair. 

We look for a signal in the scatter plots of the invariant mass of the $\eta$ candidate vs.  the missing mass $M_{miss}=\sqrt{(M_{\Utwo}-E_{\eta})^2-p_{\eta}^2}$ which are presented in Figure~\ref{figs:etascatdtel} for the $ee$ channel and in Figure~\ref{figs:etascatdtmu} for the $\mu\mu$ channel (the boxes denote our signal regions which are optimized using a Monte Carlo simulation). In Table~\ref{tab:etaevents} we list the number of observed events for the decay channels under consideration along with the detection efficiencies of each individual channel determined from Monte Carlo simulation.

\begin{table}[h]
\center
\caption[]{\small Numbers of observed events and efficiencies  for the \decthree,$\Uone\rightarrow l^+l^-$ measurement in different $\eta$ decay modes.}\label{tab:etaevents}
\begin{tabular}{lccccc}
  &   & \multicolumn{2}{c}{$ee$ channel} & \multicolumn{2}{c}{$\mu\mu$ channel} \\ \cline{3-6}
Decay mode & BR & $N^{observed}$ & Efficiency (\%) & $N^{observed}$ & Efficiency (\%) \\ \hline
 $\eta\rightarrow 3\pi^0$ & 0.319 &  0 &$  2.4\pm  0.3$&  0 &$  2.3\pm  0.3$ \\
 $\eta\rightarrow 2\gamma$ & 0.389 & 13 &$ 38.4\pm  1.6$&  3 &$ 46.9\pm  1.9$ \\
 $\eta\rightarrow \pi^+\pi^-\pi^0$ & 0.236 &  0 &$  8.9\pm  0.8$&  0 &$ 10.5\pm  0.9$ \\
 $\eta\rightarrow \pi^+\pi^-\gamma$ & 0.049 &  1 &$ 17.5\pm  2.0$&  2 &$ 22.0\pm  2.2$\\
\end{tabular}
\end{table}

To convert the numbers from Table~\ref{tab:etaevents} into branching ratios or upper limits we have to consider the sources of possible background contamination. The \decone\ transition with initial or final state radiation can mask the \decthree\ transition with \dectwel\ and the transition \dectwo\ where two photons from different $\pi^0$'s escape detection can mask the $\eta$ transition with \decnine. To estimate these two backgrounds, we used our Monte Carlo sample of the exclusive dipion transitions with 50,000 events in the charged mode and 40,000 events in the neutral mode which we subject to the $\eta$ transition selection criteria. After scaling, we found the background to be \Neebkgpizerotwo(\Nmmbkgpizerotwo) events in the \dectwel\, $ee(\mu\mu$) channel and \Neebkgpifour(\Nmmbkgpifour) events in the \decnine\, $ee(\mu\mu$) channel. We did not observe any background events in the \decten\ or \decelev\ channels. Another source of possible background are the cascade radiative decays $\Utwo \rightarrow \gamma\chi_{b}\rightarrow \gamma\gamma\Uone$. This contamination was estimated based on a 15,000 Monte Carlo sample of the cascade radiative decays. We found no background events from this source. To estimate the background from radiative QED and other possible nonresonant processes we used a data sample corresponding to an integrated luminosity of \Lumtwosoff\ pb$^{-1}$ of $e^+e^-$ annihilations taken at $\sqrt{s}=9.98$ GeV, just below the \Utwo\ resonance. After scaling for luminosity and energy differences we found \Neebkgconttwo(\Nmmbkgconttwo) background events for the \decnine\ mode in the $ee(\mu\mu)$ channel and no background events for the three remaining $\eta$ decay modes. The results of the background study are summarized in Table~\ref{tab:etabackground}.     

\begin{table}[h]
\center
\caption[]{\small Number of expected background events for the \decthree,$\Uone\rightarrow l^+l^-$ transition in different $\eta$ decay modes for our \Utwo\ resonance data sample.}\label{tab:etabackground}
\begin{tabular}{lccccc}
    & \multicolumn{4}{c}{Sources of background, events in $ee(\mu\mu)$ channel} &  \\ \cline{2-5}
    Decay mode     & $\pi^+\pi^-$ & $\pi^0\pi^0$  & $\gamma\gamma$ cascade & \Utwo, continuum & Total \\ \hline
 $\eta\rightarrow 3\pi^0$ &  0(0) &  0(0) &  0(0) &  0(0) &  0(0) \\
 $\eta\rightarrow 2\gamma$ &  0(0) &  0.2(0.2) &  0(0) & 14.2(0) & 14.5(0.2) \\
 $\eta\rightarrow\pi^+\pi^-\pi^0$& 0(0)& 0(0)& 0(0)& 0(0) & 0(0)\\
 $\eta\rightarrow\pi^+\pi^-\gamma$& 0.3(0.6)& 0(0)& 0(0)& 0(0)& 0.3(0.6)\\
\end{tabular}
\end{table}

Although the above study shows that in the $\mu\mu$ channel the expected number of events from background processes in the signal region is not consistent with the number of observed events, some of the signal events lie very close to the signal box boundary which leads us to interpret our signal candidates as smearing of background events into the signal region. Therefore we (conservatively) do not calculate a branching ratio but set an upper limit. Because the mode \decnine\ in the $ee$ channel is so ``noisy'' we exclude it from the procedure of setting the upper limit on the branching ratio. 
After taking into account the errors on efficiencies, on the \Uone\ leptonic and $\eta$ branching ratios, we set the following upper limit:\footnote{To calculate an upper limit on the number of signal events we follow the procedure suggested by PDG~\cite{poisson} and include the systematic errors according to~\cite{syserinuplim}.}
\[ {\cal B}(\decthree) < \Nthre\ (90\% C.L.)\]
The results from other experiments are given in Table~\ref{tab:etacomp}. 

\begin{table}[h]
\center
\caption[]{\small Upper limits on ${\cal B}(\decthree)\ (90\% C.L.$).}\label{tab:etacomp}
\begin{tabular}{ll}
CLEO~\cite{CLEO84}            & $<0.010$ \\
Crystal Ball~\cite{CBALL87}   & $<0.007$ \\
ARGUS~\cite{ARGUS87}          & $<0.005$ \\
CUSB~\cite{CUSB84}            & $<0.002$ \\
this analysis              & $<\Nthre$ \\
\end{tabular}
\end{table}

In the multipole expansion of the gluon color field, $\pi\pi$ transitions proceed via $E1\cdot E1$ emission but the lowest order transition allowed by the quantum numbers of the $\eta$-meson is $E1\cdot M2$ or $M1\cdot M1$ emission. This results in a suppression of the $\eta$ transition compared to the $\pi^+\pi^-$ transition by a factor of $\approx 5\cdot 10^{-3}$ ~\cite{Vol-Zai}, so the branching ratio for \decthree\ is expected to be around 0.001, below the current upper limit. Since for the chromomagnetic transitions the transition amplitude varies as $m^{-4}_{quark}$, the ratio ${\cal B}(\Utwo\rightarrow \Uone\eta)/{\cal B}(\Utwo\rightarrow \Uone\pi^+\pi^-)$ should be substantially smaller than the ratio ${\cal B}(\psi(3685)\rightarrow \psi\eta)/{\cal B}(\psi(3685)\rightarrow \psi\pi^+\pi^-)=0.083$. Yan~\cite{Yan} obtained the formula:
\[r_{b/c}=\frac{\Gamma(\decthree)}{\Gamma(\psi(3685)\rightarrow \psi\eta)}\simeq \left(\frac{m_{c}}{m_{b}}\right)^{4} \left(\frac{p_{\Upsilon}}{p_{\psi}}\right)^{3}\simeq \frac{1}{275}\]
where $p_{\Upsilon}$ and $p_{\psi}$ are the decay momenta. Our experimental value is $r_{b/c} < 1/61$, using $\Gamma_{tot}(\psi(3685))=277$ keV and ${\cal B}(\psi(3685)\rightarrow \psi\eta)=0.027$; this is 15 times smaller than the suppression expected from phase space alone (a factor of four). Our results are clearly consistent with the multipole expansion formalism.

\section{Transition \boldmath{\decfour}}

We also studied the isospin violating transition \decfour\ with \deceigh\ and $\pi^0\rightarrow \gamma\gamma$. The same set of selection criteria as in the exclusive \decpiall\ study is applied to lepton candidates and the same set of selection criteria on photons that was used for direct reconstruction of $\eta$'s from two $\gamma$'s in the \decthree\  study is applied here. Additional requirements are: (1) $p_{\pi^0} < 0.6$ GeV/$c$, (2) the number of good charged tracks equals two, (3) the number of showers unmatched to tracks less than three, (4) the cosine of the angle between the $\pi^0$ and the dilepton system $cos\theta_{\pi ll} < -0.9$ (to reduce the background from QED processes), and (5) 9.21 GeV/$c^2 < m_{ll} < 9.71$ GeV/$c^2$ where $m_{ll}$ is the dilepton invariant mass.

\begin{figure}[htb]
\center
\epsfig{file=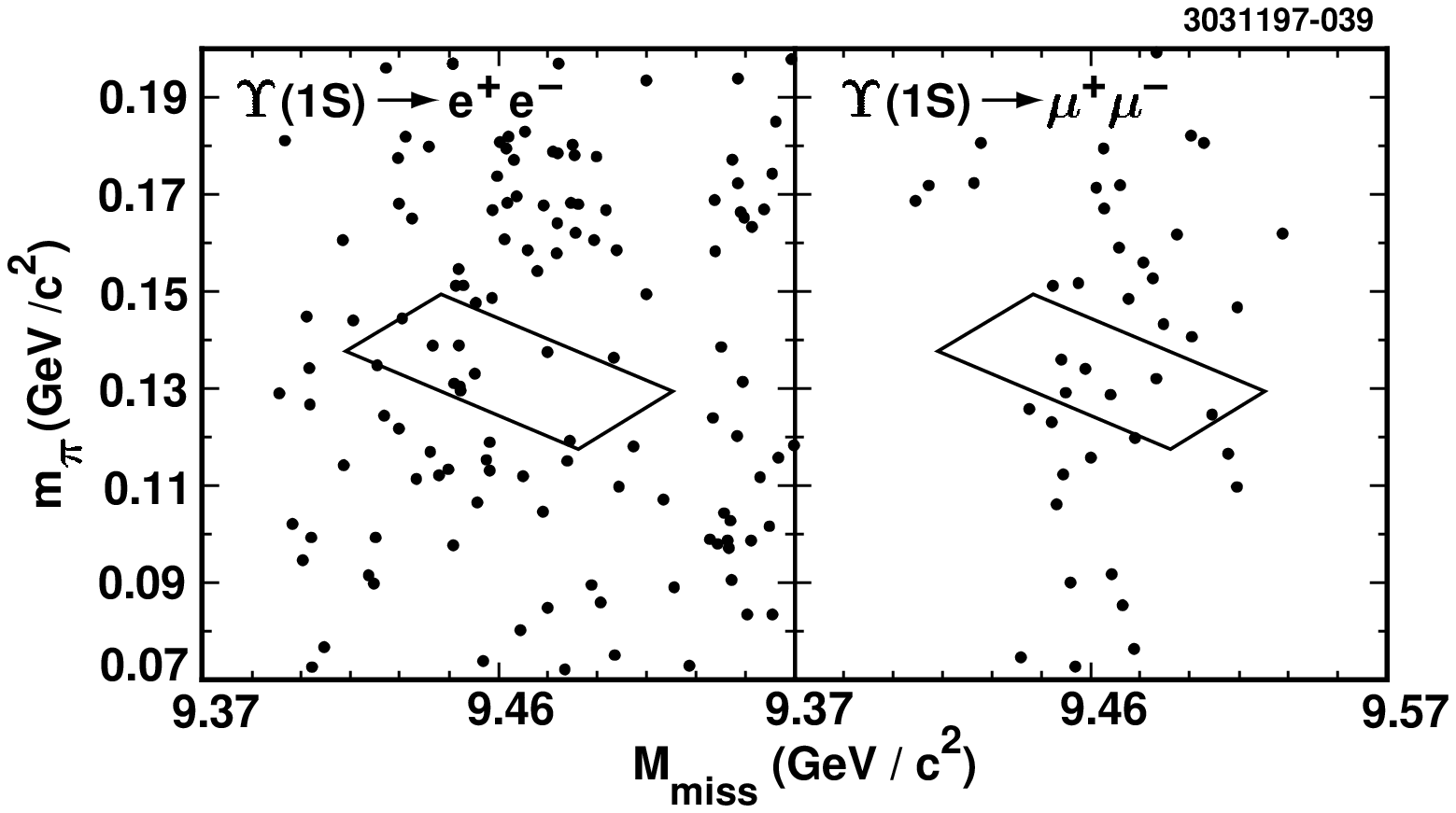, height=8cm}
\caption[]{Scatter plot of $\pi^0$ invariant mass vs. missing mass for $\Utwo \rightarrow \Uone \pi^0$, $\Uone\rightarrow l^+l^-$\ from \Utwo\ resonance data.}\label{figs:pi0scatdt}
\end{figure}

As in the search for the $\eta$ transition, we look for a signal in the scatter plot of the  $\pi^0$ invariant mass $m_{\pi^0}$ vs. the missing mass $M_{miss}=\sqrt{(M_{\Utwo}-E_{\pi^0})^2-p_{\pi^0}^2}$.  In Figure~\ref{figs:pi0scatdt} the scatter plots from the \Utwo\ resonance data sample are displayed for both dilepton channels (Monte Carlo simulation is used to optimize the signal regions denoted by the solid boxes).

\begin{table}[h]
\center
\caption[]{\small Number of observed events and efficiencies for the \decfour, $\Uone\rightarrow l^+l^-$ transition.}\label{tab:pi0data}
\begin{tabular}{ccc}
    & $N^{observed}$ & Efficiency (\%) \\ \hline
 $ee$ &  9 & $ 29.3\pm  0.8 $ \\
 $\mu\mu$ &  6 & $ 36.3\pm  0.9 $ \\ 
\end{tabular}
\end{table}

Within the signal region, we find \Npizeltot\ events in the $ee$ channel and \Npizmutot\ events in the $\mu\mu$ channel. The efficiencies, which are based on Monte Carlo simulation, are given in Table~\ref{tab:pi0data}.

We used a ``grand side-band'' technique, to estimate the background: we counted the events in the ``grand side-band'' (in Figure~\ref{figs:pi0scatdt} it is all the area outside the signal box for the $ee$ channel and a vertical strip between 9.41 GeV and 9.51 GeV in $M_{miss}$, excluding the signal box, for the $\mu\mu$ channel) and extrapolated the background event yield into the signal region. The results are given in Table~\ref{tab:pi0gsb}.

\begin{table}[h]
\center
\caption[]{\small Numbers of the events from the ``grand side-band'' subtraction technique.}\label{tab:pi0gsb}
\begin{tabular}{cccc} 
  & $N_{sideband}$ & $N_{signal-region}$ & Estimated $N_{signal-region}^{background}$ \\ \hline 
 $ee$ & 130 &   9 &  8.4 \\
 $\mu\mu$ &  37 &   6 &  4.5 \\
\end{tabular}
\end{table}

As seen in the table, using the ``grand side-band'' subtraction technique we expect \Npizbkgtotgsb background events compared to the total of 15 observed events. This corresponds to an upper limit:
\[ {\cal B}(\decfour) < \Brpizgsb\  (90\%C.L.) \]
This is the lowest upper limit on the $\pi^0$ transition to date. The only other experiment that studied this transition was Crystal Ball (Table~\ref{tab:pi0comp}).

\begin{table}[ht]
\center
\caption[]{\small Upper limits on ${\cal B}(\decfour)\ (90\% C.L.$).}\label{tab:pi0comp}
\begin{tabular}{ll}
Crystal Ball~\cite{CBALL87}   & $<0.008$ \\
this analysis              & $<\Brpizgsb$ \\
\end{tabular}
\end{table}

The \decfour\ transition can occur because of a breaking of the isotopic symmetry due to the mass difference between the $u$ and $d$ quarks, and its rate is expected to be lower than the \decthree\ rate. In the context of the multipole expansion, this ratio is given by~\cite{Vol-Zai}:
\[r_{\pi^0/\eta}\simeq \frac{\Gamma( (2S)\rightarrow (1S)\pi^0)}{\Gamma((2S)\rightarrow (1S)\eta)}=3\left(\frac{m_{d}-m_{u}}{m_{d}+m_{u}}\right)^2 \left(\frac{m_{\pi}}{m_{\eta}}\right)^{4} \left(\frac{p_{\pi}}{p_{\eta}}\right)^{3}\]
With $(m_{d}-m_{u})/(m_{d}+m_{u})\approx 0.3$~\cite{udratio} this gives $r_{\pi^0/\eta}\approx 0.022$ for charmonium which is in reasonable agreement with the experimental value of 0.037. For bottomonium we have $r_{\pi^0/\eta}\approx 0.14$ and $\Gamma(\decfour)\approx 0.003$ keV (using $\Gamma(\decthree)=0.02$ keV from Kuang-Yan~\cite{Kuang-Yan}) which is more than an order of magnitude below our upper limit of 0.048 keV.

\section{Summary}

We have measured various experimental quantities for the hadronic transitions from the \Utwo\ to \Uone\ including branching ratios, the dipion invariant mass spectra, and angular distributions. Using the PDG value for the full width of the \Utwo\ resonance $\Gamma = 44$ keV~\cite{ufourwidth}, we also calculate the partial widths for the corresponding transitions. Table~\ref{tab:summary} reports our measurements of the branching ratios and partial widths compared with previous world averages and theoretical calculations by Kuang and Yan~\cite{Kuang-Yan}. Our results are consistent with previous experiments as well as theoretical predictions. We determine an upper limit on the branching ratio of \decthree\ and set a new upper limit on the branching ratio of the \decfour\ transition.

\begin{table}[ht]
\center
\caption[]{\small Summary of the branching ratios and rates of hadronic transitions of \Utwo.}\label{tab:summary}
\begin{tabular}{lcccc} 
                   & \multicolumn{2}{c}{Branching ratio} & \multicolumn{2}{c}{Rates (keV)} \\  
        Decay & Experiment & World Avg. & Experiment & Kuang-Yan\\ \hline
\decone       &  $\Broneave^{a}$  & $0.185\pm 0.008$ & $8.4\pm 0.5$ & 8.8 \\
\dectwo       &  \BRtwo  & $0.088\pm 0.011$ & $4.0\pm 0.4$ & 4.4 \\
\decthree          &$<\Nthre$ & $<0.002$         & $<0.12$ & 0.02 \\
\decfour           &$<\Brpizgsb$ & $<0.008$         & $<0.048$ & 0.003 \\
\end{tabular}
\raggedright {\small $^a$ average over the exclusive and inclusive measurements.}
\end{table}

We also calculate the leptonic branching ratios of the \Uone:\  $B_{ee}={\cal B}(\Uone\rightarrow e^+e^-)=\Nsix$\ and $B_{\mu\mu}={\cal B}(\Uone\rightarrow \mu^+\mu^-)=\Nsevn$\ which are in good agreement with PDG values.

The dipion invariant mass spectrum we observe in \decpiall\ transitions is well described by both the model of multipole expansion of the gluon color field~\cite{Zhou-Kuang} and the model of Novikov, Shifman, Voloshin and Zakharov who used the general form of the QCD field tensor $G^{a}_{\mu\nu}$\ to obtain a hadronization matrix element in the chiral limit~\cite{Vol-Zakh,Nov-Shif}.

The angular distributions of the final state particles in \decpiall\ show a strong $s-$wave dominance, as expected from theory. A small $d-$wave contribution  on the order of 4\% may be present in our data.\\

\centerline{\bf ACKNOWLEDGEMENTS}
\smallskip
We gratefully acknowledge the effort of the CESR staff in providing us with
excellent luminosity and running conditions.
J.P.A., J.R.P., and I.P.J.S. thank                                           
the NYI program of the NSF, 
M.S. thanks the PFF program of the NSF,
G.E. thanks the Heisenberg Foundation, 
%
%
K.K.G., M.S., H.N.N., T.S., and H.Y. thank the
OJI program of DOE, 
J.R.P., K.H., M.S. and V.S. thank the A.P. Sloan Foundation,
R.W. thanks the 
Alexander von Humboldt Stiftung, 
M.S. thanks Research Corporation, 
and S.D. thanks the Swiss National Science Foundation 
for support.
This work was supported by the National Science Foundation, the
U.S. Department of Energy, and the Natural Sciences and Engineering Research 
Council of Canada.


\begin{thebibliography}{99}
\bibitem{numberofutwos} To calculate the total number of \Utwo\ resonant decays, we use the formula $N_{\Utwo}=N_{had}/\epsilon_{had}=(N^{on}-N^{off}(L^{on}s^{off}/L^{off}s^{on}))/\epsilon_{had}$, where $N^{on,off}$ is the number of observed hadronic events, $L^{on,off}$ is the integrated luminosity on- and off-resonance, $\sqrt{s}=E_{cm}$, and $\epsilon_{had}$ is the probability for any $\Upsilon(2S)$ event to pass our hadronic event selection criteria (based on Monte Carlo).
\bibitem{Yanmodified} This particular form is derived from a formula in~\cite{Yan} and explicitly quoted in~\cite{CLEO84}.
\end{thebibliography}
\end{document}